\documentclass[preprint,prd,aps,showpacs,showkeys,nofootinbib]{revtex4}
\usepackage{graphicx}
\usepackage{makecell}
\usepackage{bm}
\usepackage{url}
\begin{document}

Published in JHEP 05 (2023) 134

\title{Lightest Higgs boson decays $h\rightarrow MZ$ in the $\mu$ from $\nu$ supersymmetric standard model}

\author{Chang-Xin Liu$^{a,b,c}$\footnote{LIUchangxinZ@163.com},
Hai-Bin Zhang$^{a,b,c}$\footnote{Corresponding author: hbzhang@hbu.edu.cn},\\
Jin-Lei Yang$^{a,b,c}$\footnote{JLYangJL@163.com},
Shu-Min Zhao$^{a,b,c}$\footnote{zhaosm@hbu.edu.cn},
Tai-Fu Feng$^{a,b,c,d,e}$\footnote{Corresponding author: fengtf@hbu.edu.cn}}

\affiliation{$^a$Department of Physics, Hebei University, Baoding, 071002, China\\
$^b$Key Laboratory of High-precision Computation and Application of Quantum Field Theory of Hebei Province, Baoding, 071002, China\\
$^c$Research Center for Computational Physics of Hebei Province, Baoding, 071002, China\\
$^d$Department of Physics, Guangxi University, Nanning, 530004, China\\
$^e$College of Physics, Chongqing University, Chongqing, 400044, China}

\begin{abstract}
We study the lightest Higgs boson decays $h\rightarrow MZ$ in the $\mu$ from $\nu$ supersymmetric standard model ($\mu\nu$SSM), where $M$ is a vector meson $(\rho,\omega,\phi,J/\Psi,\Upsilon)$. Compared to the minimal supersymmetric standard model (MSSM), the $\mu\nu$SSM introduces three right-handed neutrino superfields, which lead to the mixing of the Higgs doublets with the right-handed sneutrinos. The mixing affects the lightest Higgs boson mass and the Higgs couplings. In suitable parameter space, the $\mu\nu$SSM can give large new physics (NP) contributions to the signal strengths of $h\rightarrow MZ$ and $h\rightarrow \gamma\gamma$, which may be detected by a 100 TeV collider or the other future high energy colliders.

\end{abstract}


\maketitle

\section{Introduction\label{sec1}}

Since the Higgs boson was discovered by the Large Hadron Collider (LHC) \cite{mh-ATLAS,mh-CMS}, the measured mass of the Higgs boson now is~\cite{PDG1}
\begin{eqnarray}
m_h=125.25\pm 0.17\: {\rm{GeV}}.
\label{mh-exp}
\end{eqnarray}

Therefore, the accurate Higgs boson mass gives most stringent constraint on parameter space of various extensions of the standard model (SM). The SM Higgs boson mass in the tree level can be written as $m_{h}=2\sqrt{\lambda_{SM}}\upsilon$ \cite{PDG1}, where $\lambda_{SM}$ is the self coupling parameter. $\upsilon=(2\sqrt{2}G_{F})^{-1/2}\approx$ 174 GeV is the expectation value of the Higgs field, with $G_{F}$ is the Fermi coupling. Based on the experimental measurement of the Higgs boson mass, we can get $\lambda_{SM}\simeq$ 0.13.

The next step is focusing on searching for the properties of the Higgs boson both experimentally and theoretically. Most Higgs couplings are well established. Active searches are going on the di-Higgs production to probe Higgs self-coupling unambiguously. Search for NP in couplings is still very motivated. Any deviation in the measurement of the couplings of the Higgs boson will be a signal beyond SM.

As one of the extensions of the SM, the $\mu$ from $\nu$ supersymmetric standard model ($\mu$$\nu$SSM) \cite{mnSSM,mnSSM1,mnSSM1-1,parameter-space,mnSSM2,mnSSM2-1,Zhang1,Zhang2} can solve the $\mu$ problem~\cite{m-problem} of the minimal supersymmetric standard model (MSSM)~\cite{MSSM,MSSM1,MSSM2,MSSM3,MSSM4}, through introducing three singlet right-handed neutrino superfields $\hat{\nu}_i^c$ ($i=1,2,3$). The neutrino superfields lead the mixing of the neutral components of the Higgs doublets with the right-handed sneutrinos, that is different from the Higgs sector of the MSSM. The mixing can change the Higgs couplings, and influence the decay processes of the Higgs bosons.

In our previous work, the 125 GeV SM-like Higgs boson decay modes $h\rightarrow\gamma\gamma$, $h\rightarrow VV^*$ ($V=Z,W$), $h\rightarrow f\bar{f}$ ($f=b,\tau$), $h\rightarrow \mu\tau$, $h\rightarrow Z\gamma$, the masses of the Higgs bosons in the $\mu\nu$SSM have been researched~\cite{HZrr,muon,MASS,HZr}. In this work, we study the 125 GeV SM-like Higgs boson rare decays $h\rightarrow MZ$ in the framework of the $\mu\nu$SSM, where $M$ is a vector meson ($\rho, \omega, \phi, J/\psi, \Upsilon$).
For the processes $h\rightarrow MZ$~\cite{HVZ-sm,HVZ-zhao,HVZ-NP1,HVZ-sm1,HVZ-sm2,HVZ-sm3}, there are two types of decay topologies: one is the direct contributions which the SM-like Higgs boson couples to the quarks forming the meson, and another is the indirect contributions resulting from a $h\rightarrow \gamma^* Z$ transition followed by the conversion of the off-shell boson into a vector meson. Compared to the SM-like Higgs boson coupling to the quarks in the direct contributions, here the effective $h\gamma Z$ vertex in the indirect contributions will give more NP contributions. So the indirect contributions of the decays $h\rightarrow MZ$ are more obvious than the direct contributions to search for NP. In this work, the QCD factorization \cite{QCD1,QCD1-1,QCD2,QCD3} is used for the rare SM-like Higgs boson decays $h\rightarrow MZ$.

The paper is organized as follows. In Sec.~\ref{sec2}, we introduce the $\mu\nu$SSM briefly, about the superpotential and the soft SUSY-breaking terms. In Sec.~\ref{sec3}, we study the decay processes $h\rightarrow MZ$  in the $\mu\nu$SSM. Sec.~\ref{sec4} and Sec.~\ref{sec5}, we show the numerical analysis and the conclusion.

\section{the $\mu\nu$SSM\label{sec2}}

In addition to the MSSM Yukawa couplings for quarks and charged leptons, the superpotential of the $\mu\nu$SSM contains Yukawa couplings for neutrinos, two additional types of terms involving the Higgs doublet superfields $\hat H_u$ and $\hat H_d$, and the right-handed neutrino superfields  $\hat{\nu}_i^c$,~\cite{mnSSM}
\begin{eqnarray}
&&W={\epsilon _{ab}}\left( {Y_{{u_{ij}}}}\hat H_u^b\hat Q_i^a\hat u_j^c + {Y_{{d_{ij}}}}\hat H_d^a\hat Q_i^b\hat d_j^c
+ {Y_{{e_{ij}}}}\hat H_d^a\hat L_i^b\hat e_j^c \right)  \nonumber\\
&&\hspace{0.95cm}
+ {\epsilon _{ab}}{Y_{{\nu _{ij}}}}\hat H_u^b\hat L_i^a\hat \nu _j^c -  {\epsilon _{ab}}{\lambda _i}\hat \nu _i^c\hat H_d^a\hat H_u^b + \frac{1}{3}{\kappa _{ijk}}\hat \nu _i^c\hat \nu _j^c\hat \nu _k^c ,
\label{eq-W}
\end{eqnarray}
where $\hat H_u^T = \Big( {\hat H_u^ + ,\hat H_u^0} \Big)$, $\hat H_d^T = \Big( {\hat H_d^0,\hat H_d^ - } \Big)$, $\hat Q_i^T = \Big( {{{\hat u}_i},{{\hat d}_i}} \Big)$, $\hat L_i^T = \Big( {{{\hat \nu}_i},{{\hat e}_i}} \Big)$ (the index $T$ denotes the transposition) represent $SU(2)$ doublet superfields, and $\hat u_i^c$, $\hat d_i^c$, and $\hat e_i^c$ are the singlet up-type quark, down-type quark and charged lepton superfields, respectively.  In addition, $Y_{u,d,e,\nu}$, $\lambda$, and $\kappa$ are dimensionless matrices, a vector, and a totally symmetric tensor.  $a,b=1,2$ are SU(2) indices with antisymmetric tensor $\epsilon_{12}=1$, and $i,j,k=1,2,3$ are generation indices. The summation convention is implied on repeated indices in the following.

In the superpotential, if the scalar potential is such that nonzero vacuum expectative values (VEVs) of the scalar components ($\tilde \nu _i^c$) of the singlet neutrino superfields $\hat{\nu}_i^c$ are induced, the effective bilinear terms $\epsilon _{ab} \varepsilon_i \hat H_u^b\hat L_i^a$ and $\epsilon _{ab} \mu \hat H_d^a\hat H_u^b$ are generated, with $\varepsilon_i= Y_{\nu _{ij}} \left\langle {\tilde \nu _j^c} \right\rangle$ and $\mu  = {\lambda _i}\left\langle {\tilde \nu _i^c} \right\rangle$,  once the electroweak symmetry is broken. The last term Eq.~(\ref{eq-W})  generates the effective Majorana masses for neutrinos at the electroweak scale. Therefore, the $\mu\nu$SSM can generate three tiny neutrino masses at the tree level through TeV scale seesaw mechanism
\cite{mnSSM,neutrino-mass,neu-mass1,neu-mass2,neu-mass3,neu-mass4,neu-mass5,neu-mass6}.

It is worth explaining why TeV scale  seesaw was chosen. Through a seesaw on the scale of the Grand Unified Theory (GUT), one can get the Yukawa couplings of order one for neutrinos. But we know that the Yukawa coupling of the electron is on the order of $10^{-6}$, and the Yukawa couplings of neutrinos can also be around on the order of $10^{-6}$ instead of one. In the TeV scale seesaw, this is sufficient to reproduce the neutrino mass, if the Yukawa coupling of the neutrino is of the same order as the Yukawa coupling of the electron \cite{mnSSM}. Here it is important to note that the VEVs of the left-handed sneutrinos $\upsilon_{\nu_i}$ are generally small.
We know that the Dirac masses for the neutrinos $m_{D_i}=Y_{\upsilon_{\nu_i}}\upsilon_{u}\lesssim 10^{-4}$ GeV in the TeV scale seesaw. So we can get an estimate of the VEVs of the left-handed sneutrinos, $\upsilon_{\nu_i}\lesssim m_{D_i}\lesssim 10^{-4}$ GeV, which means that $\upsilon_{\nu_i}\ll\upsilon_{d},\upsilon_{u}$ \cite{mnSSM,mnSSM1}.

In supersymmetric (SUSY) extensions of the SM, the R-parity of a particle is defined as $R = (-1)^{L+3B+2S}$~\cite{MSSM,MSSM1,MSSM2,MSSM3,MSSM4}. R-parity is violated if either the baryon number ($B$) or lepton number ($L$) is not conserved, where $S$ denotes the spin of concerned component field. The last two terms in Eq.~(\ref{eq-W}) explicitly violate lepton number and R-parity. For example, if one assigns $L = 1$ to the right-handed neutrino superfields, then the last term ${1\over3}\kappa_{ijk}\hat{\nu}_{i}^{c}\hat{\nu}_{j}^{c}\hat{\nu}_{k}^{c}$ of Eq. (2) violates the lepton number by three units contrary to the $\hat{L}_{i}\hat{L}_{j}\hat{e}^{c}_{k}$ term of the R-parity violating MSSM which shows $\Delta L = 1$ effect. R-parity breaking implies that the lightest supersymmetric particle (LSP) is no longer stable. In this context, the neutralino or the left-handed and right-handed sneutrino are no longer candidates for the dark matter (DM). However, other SUSY particles such as the gravitino or the axino can still be used as dark matter candidates~\cite{mnSSM1,mnSSM1-1,neu-mass3,DM1,DM2,DM3,DM4,DM5,DM6}.

The general soft SUSY-breaking terms of the $\mu\nu$SSM are given by
\begin{eqnarray}
&&- \mathcal{L}_{soft}=m_{{{\tilde Q}_{ij}}}^{\rm{2}}\tilde Q{_i^{a\ast}}\tilde Q_j^a
+ m_{\tilde u_{ij}^c}^{\rm{2}}\tilde u{_i^{c\ast}}\tilde u_j^c + m_{\tilde d_{ij}^c}^2\tilde d{_i^{c\ast}}\tilde d_j^c
+ m_{{{\tilde L}_{ij}}}^2\tilde L_i^{a\ast}\tilde L_j^a  \nonumber\\
&&\hspace{1.7cm} +  m_{\tilde e_{ij}^c}^2\tilde e{_i^{c\ast}}\tilde e_j^c + m_{{H_d}}^{\rm{2}} H_d^{a\ast} H_d^a
+ m_{{H_u}}^2H{_u^{a\ast}}H_u^a + m_{\tilde \nu_{ij}^c}^2\tilde \nu{_i^{c\ast}}\tilde \nu_j^c \nonumber\\
&&\hspace{1.7cm}  +  \epsilon_{ab}{\left[{{({A_u}{Y_u})}_{ij}}H_u^b\tilde Q_i^a\tilde u_j^c
+ {{({A_d}{Y_d})}_{ij}}H_d^a\tilde Q_i^b\tilde d_j^c + {{({A_e}{Y_e})}_{ij}}H_d^a\tilde L_i^b\tilde e_j^c + {\rm{H.c.}} \right]} \nonumber\\
&&\hspace{1.7cm}  + \left[ {\epsilon _{ab}}{{({A_\nu}{Y_\nu})}_{ij}}H_u^b\tilde L_i^a\tilde \nu_j^c
- {\epsilon _{ab}}{{({A_\lambda }\lambda )}_i}\tilde \nu_i^c H_d^a H_u^b
+ \frac{1}{3}{{({A_\kappa }\kappa )}_{ijk}}\tilde \nu_i^c\tilde \nu_j^c\tilde \nu_k^c + {\rm{H.c.}} \right] \nonumber\\
&&\hspace{1.7cm}  -  \frac{1}{2}\left({M_3}{{\tilde \lambda }_3}{{\tilde \lambda }_3}
+ {M_2}{{\tilde \lambda }_2}{{\tilde \lambda }_2} + {M_1}{{\tilde \lambda }_1}{{\tilde \lambda }_1} + {\rm{H.c.}} \right).
\end{eqnarray}
Here, the first two lines contain mass squared terms of squarks, sleptons, and Higgses. The next two lines consist of the trilinear scalar couplings. In the last line, $M_3$, $M_2$, and $M_1$ denote Majorana masses corresponding to $SU(3)$, $SU(2)$, and $U(1)$ gauginos $\tilde{\lambda}_3$, $\tilde{\lambda}_2$, and $\tilde{\lambda}_1$, respectively. In addition to the terms from $\mathcal{L}_{soft}$, the tree-level scalar potential receives the usual $D$- and $F$-term contributions~\cite{mnSSM1,mnSSM1-1}.

Once the electroweak symmetry is spontaneously broken, the neutral scalars develop in general the VEVs:
\begin{eqnarray}
\langle H_d^0 \rangle = \upsilon_d , \qquad \langle H_u^0 \rangle = \upsilon_u , \qquad
\langle \tilde \nu_i \rangle = \upsilon_{\nu_i} , \qquad \langle \tilde \nu_i^c \rangle = \upsilon_{\nu_i^c} .
\end{eqnarray}
One can define the neutral scalars as
\begin{eqnarray}
&&H_d^0=\frac{h_d + i P_d}{\sqrt{2}} + \upsilon_d, \qquad\; \tilde \nu_i = \frac{(\tilde \nu_i)^\Re + i (\tilde \nu_i)^\Im}{\sqrt{2}} + \upsilon_{\nu_i},  \nonumber\\
&&H_u^0=\frac{h_u + i P_u}{\sqrt{2}} + \upsilon_u, \qquad \tilde \nu_i^c = \frac{(\tilde \nu_i^c)^\Re + i (\tilde \nu_i^c)^\Im}{\sqrt{2}} + \upsilon_{\nu_i^c},
\end{eqnarray}
and
\begin{eqnarray}
\tan\beta={\upsilon_u\over{\sqrt{\upsilon_d^2+\upsilon_{\nu_i}\upsilon_{\nu_i}}}}
\approx {\upsilon_u\over\upsilon_d},
\qquad\upsilon=\sqrt{\upsilon_{u}^2+\upsilon_{d}^2+\upsilon_{\nu_i}\upsilon_{\nu_i}}
\approx \sqrt{\upsilon_{u}^2+\upsilon_{d}^2},
\end{eqnarray}

In the $\mu\nu$SSM, the left-  \& right-handed sneutrino VEVs lead to the mixing of the neutral components of the Higgs doublets with the left- $\&$ right-handed sneutrinos producing an $8\times8$ CP-even neutral scalar mass matrix, which can be seen in Refs.~\cite{mnSSM1,mnSSM1-1,Zhang1,MASS}. The mixing gives a rich phenomenology in the Higgs sector of the $\mu\nu$SSM \cite{mnSSM1,mnSSM2,mnSSM1-1,parameter-space,mnSSM2-1,HZrr,muon,MASS,HZr}. In Ref.~\cite{MASS}, we have analyzed the radiative corrections to the Higgs boson masses in $\mu\nu$SSM, which will be used in the following numerical analysis.
The mixing can influence the mass and the coupling of the lightest Higgs boson, which can give new effect to the lightest Higgs boson weak hadronic decays $h\rightarrow MZ$.

\section{The Processes of $h\rightarrow MZ$ \label{sec3}}

\begin{figure}
\setlength{\unitlength}{1mm}
\begin{minipage}[c]{0.25\textwidth}
\centering
\includegraphics[width=1.5in]{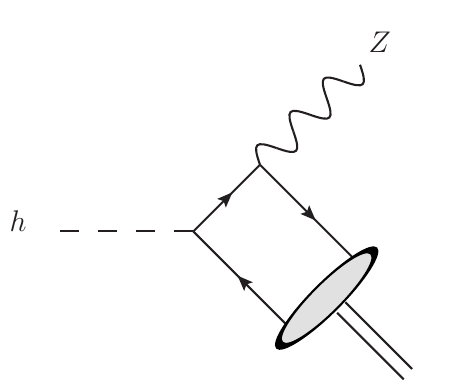}
\end{minipage}%
\begin{minipage}[c]{0.25\textwidth}
\centering
\includegraphics[width=1.5in]{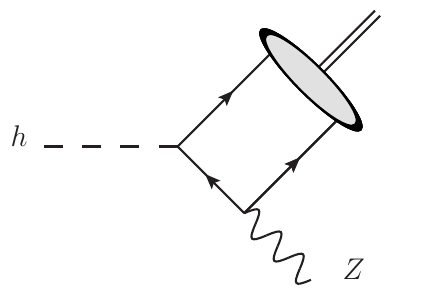}
\end{minipage}%
\begin{minipage}[c]{0.25\textwidth}
\centering
\includegraphics[width=1.5in]{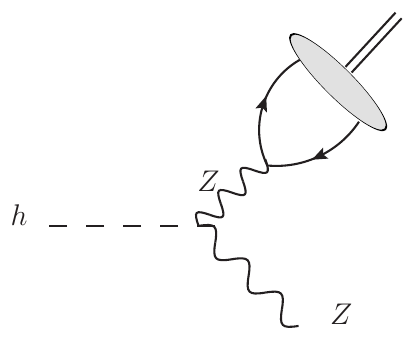}
\end{minipage}
\begin{minipage}[c]{0.24\textwidth}
\centering
\includegraphics[width=1.5in]{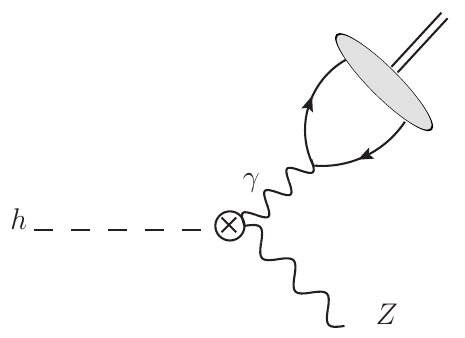}
\end{minipage}
\caption[]{The dominating Feynman diagrams for $h\rightarrow MZ$, where M is a vector meson ($\rho, \omega, \phi, J/\psi, \Upsilon$)~\cite{HVZ-sm}.}
\label{HVZ-Feynman}
\end{figure}

The lightest Higgs boson weak hadronic decays $h\rightarrow MZ$ are very interesting by the fact that the massive final-state gauge boson can be in a longitudinal polarization state~\cite{HVZ-sm}. In Fig.~\ref{HVZ-Feynman}, we show the dominating Feynman diagrams for $h\rightarrow MZ$. The first two graphs in Fig.~\ref{HVZ-Feynman} are the direct contributions $F_{direct}$, and the last two diagrams represent the indirect contributions $F_{ind}$. In the last graph, the crossed circle represents the effective vertex $h\rightarrow Z\gamma^*$ from the one loop diagrams. We notice that the $hZZ$ vertex exists at the tree-level while the $hZ\gamma$ vertex is mediated by one-loop. The $h\rightarrow Z \gamma^{\ast}$ process can be used to probe for NP, so we will focus on discussing $h\rightarrow Z \gamma^{\ast}$.

The process $h\rightarrow J/\psi Z$ has been studied experimentally \cite{experiment1}, and a search of the decays $h\rightarrow\rho Z$ and $h\rightarrow\phi Z$ has been presented in Ref. \cite{experiment2}. The upper limit on the branching ratio ${\rm{Br}}(h\rightarrow\rho Z)$ is in the range 1.04-1.31\%, or 740-940 times the SM expectation. The upper limit on ${\rm{Br}}(h\rightarrow \phi Z)$ ranges from 0.31\% to 0.40\%, or 730-950 times the SM expectation. These ranges for the upper limit depend on the polarization scenarios \cite{experiment2}. These results constitute the first experimental limits on the two decay channels. The upper limits on the branching ratios ${\rm{Br}}(h\rightarrow\rho Z)$ and ${\rm{Br}}(h\rightarrow\phi Z)$ have been set at the 95\% confidence level.

In near future, the High-Luminosity LHC (HL-LHC)~\footnote{The website of HL-LHC: \url{https://home.cern/science/accelerators/high-luminosity-lhc}.} with a peak luminosity of $7.5 \times 10^{34} \rm{cm}^{-2} \rm{s}^{-1}$, will produce at least $1.5\times 10^{7}$  SM-like Higgs bosons per year~\cite{HL-LHC}. The rare decays $h\rightarrow MZ$ with a branching fraction ${\mathcal{O}}$$(10^{-5,-6})$ in the SM or other extended models, will be difficult to be found by the HL-LHC, disturbed by background. But in the future, a potential 100 TeV collider will produce $\sim10^{10}$  SM-like Higgs bosons in the full cycle~\cite{100TeV}, which may detect the decays $h\rightarrow MZ$. For example, considered ${\rm{Br}}(h\rightarrow\rho Z)\sim 1.4\times10^{-5}$ in the SM, the decays $h\rightarrow\rho Z$ would be expected about 140 events if a rough acceptance is 0.1\%.

The proton-proton collisions has been used to search the process $h\rightarrow\rho Z$. By the CMS experiment, the final state of $h\rightarrow\rho Z$ can be a dilepton-$\pi^{+}\pi^{-}$ final state, where the $Z$ boson decays into a pair of electrons or a pair of muons, and the $\rho$ meson decays into pairs of pions \cite{experiment2}. Besides, the $\phi$ meson will decay into pairs of kaons, and the channel dilepton-$K^{+}K^{-}$ was also searched by CMS experiment \cite{experiment2}.

The decay width of $h\rightarrow MZ$ can be given as
\begin{eqnarray}
&&\Gamma(h\rightarrow MZ)={{m_h^3}\over {4\pi \upsilon^4}}\lambda^{1/2}(1,r_{Z},r_{M})(1-r_{Z}-r_{M})^2  \nonumber\\
&&\qquad\qquad\qquad\,\:\times\left[|F_{\parallel}^{MZ}|^2
+{8r_{M}r_{Z}\over(1-r_{Z}-r_{M})^2}(|F_{\perp}^{MZ}|^2+|\tilde{F}_{\perp}^{MZ}|^2)\right],
\label{decay-width}
\end{eqnarray}
here $\lambda(1,r_{Z},r_{M})=(1-r_{Z}-r_{M})^2-4r_{Z}r_{M}$, $r_{Z}={m_{Z}^2\over m_{h}^2}$, and $r_{M}={m_{M}^2\over m_{h}^2}$, $m_{M}$ is the mass of vector meson. $F_{\parallel}^{MZ}$ represent the CP-even longitudinal form factors, $F_{\perp}^{MZ}$ and $\tilde{F}_{\perp}^{MZ}$ represent the CP-even and CP-odd transverse form factors. We notice that the mass ratio $r_{M}$ is very small for all mesons, but it can make the contributions to the transverse polarization states to the $h\rightarrow MZ$ rates significant, so we still keep the mass ratio $r_{M}$ in our analysis~\cite{HVZ-sm,HVZ-zhao}.

There are two parts of the form factors of Eq.~(\ref{decay-width}), the direct and the indirect contributions. We analyse the indirect contributions firstly, which make the dominant effects to search for NP. It involves hadronic matrix elements of local currents, hence it can be calculated to all orders in QCD~\cite{HVZ-sm}. So we can obtain the indirect contributions
\begin{eqnarray}
&&F_{||\,ind}^{MZ}={\kappa_{Z}\over{1-{r_{M}/r_{Z}}}}\sum_{q}f_{M}^q\upsilon_{q}+C_{\gamma Z}{\alpha_s(m_{M})\over 4\pi}{4r_{Z}\over{1-r_{Z}-r_{M}}}\sum_{q}f_{M}^q Q_{q},
\label{form-factors1}\\
&&F_{\perp\,ind}^{MZ}={\kappa_{Z}\over{1-{r_{M}/r_{Z}}}}\sum_{q}f_{M}^q\upsilon_{q}+C_{\gamma Z}{\alpha_s(m_{M})\over 4\pi}{{1-r_{Z}-r_{M}}\over r_{M}}\sum_{q}f_{M}^q Q_{q},
\label{form-factors2}\\
&&\widetilde{F}_{\perp\,ind}^{MZ}=\widetilde{C}_{\gamma Z}{\alpha_s(m_{M})\over 4\pi}{\lambda^{1/2}(1,r_{Z},r_{M})\over r_{M}}\sum_{q}f_{M}^q Q_{q},
\label{form-factors}
\end{eqnarray}
where $\upsilon_q=T_{3}^q/2-Q_q s_{W}^2 $ are the vector couplings of the $Z$ boson to the quark $q$, $T_{3}^{q}$ and $Q_{q}$ represent the weak isospin and charge of quark $q$, and $s_W=\sin\theta_W$ with $\theta_W$ denoting the Weinberg angle. $\kappa_{Z}$ is defined as the ratio of the coupling of the SM-like Higgs boson to $Z$ boson to the corresponding SM value~\cite{PDG1}. $\alpha_s$ is the strong coupling constant \cite{PDG1,QCD1}. The flavor-specific decay constants $f_M^q$ are defined in terms of the local matrix elements~\cite{HVZ-sm,HVZ-zhao,HVZ-sm3}
\begin{eqnarray}
\langle M(k,\varepsilon)|\bar{q}\,\gamma^\mu q|0\rangle=-i f_{M}^q m_{V} \varepsilon^{*\mu}.
\end{eqnarray}
We use the relations to simplify our calculation
\begin{eqnarray}
\sum_{q}f_{M}^{q}Q_{q}=f_{M}Q_{M},\qquad\qquad\sum_{q}f_{M}^{q}\upsilon_{q}=f_{M}\upsilon_{M}.
\label{qf}
\end{eqnarray}
The mesons decay constants $f_{M}, Q_{M}, \upsilon_{M}$ for the vector meson $M=(\rho,\omega,\phi,J/\Psi,\Upsilon)$ can be seen in Table~\ref{tab1}.

\begin{table}
\begin{tabular}{|cccccc|}
\hline
Mesons $M$ &\quad$m_{M}$/GeV&\quad$f_{M}$/GeV&\quad$Q_{M}$&\quad$\upsilon_{M}$&\quad$f_{M}^{\bot}/f_{M}=f_{M}^{q\bot}/f_{M}^{q}$\\
\hline
$\rho$&0.77&0.216&$1\over{\sqrt{2}}$&$1\over{\sqrt{2}}$$({1\over{2}}-s_W^2)$&0.72\\
$\omega$&0.782&0.194&$1\over{3\sqrt{2}}$&$-{{s_{W}^2}\over{3\sqrt{2}}}$&0.71\\
$\phi$&1.02&0.223&$-{1\over{3}}$&${-{1\over4}}+{{s_{W}^2}\over3}$&0.76\\
$J/\psi$&3.097&0.403&${2\over3}$&${1\over4}-{2{s_{W}^2}\over3}$&0.91\\
$\Upsilon$&9.46&0.648&${-{1\over3}}$&${-{1\over4}}+{{s_{W}^2}\over3}$&1.09\\
\hline
\end{tabular}
\caption{The mesons decay constants $f_{M}, Q_{M}, \upsilon_{M}$ will be used in the numerical analysis, $f_{M}^{\perp}$ and $f_{M}^{q\perp}$ represent the transverse decay constants and the flavor-specific transverse decay constants \cite{HVZ-sm3}.}
\label{tab1}
\end{table}

The concrete forms of $C_{\gamma Z}$ and $\widetilde{C}_{\gamma Z}$ in Eqs.~(\ref{form-factors1}-\ref{form-factors}) are given by~\cite{HVZ-hzr,HVZ-sm,HVZ-zhao}
\begin{eqnarray}
C_{\gamma Z}=C_{\gamma Z}^{SM}+C_{\gamma Z}^{NP},\qquad\qquad \widetilde{C}_{\gamma Z}=\widetilde{C}_{\gamma Z}^{SM}+\widetilde{C}_{\gamma Z}^{NP},
\end{eqnarray}
\begin{eqnarray}
&&C_{\gamma Z}^{SM}=\sum_{q}{2N_cQ_q\upsilon_q\over 3}A_f(\tau_q,r_Z)+\sum_{l}{2Q_l\upsilon_l\over 3}A_f(\tau_l,r_Z)-{1\over2}A_{W}^{\gamma Z}(\tau_{W},r_Z), \\
&&\widetilde{C}_{\gamma Z}^{SM}=\sum_{q}\tilde{\kappa}_qN_cQ_q\upsilon_q B_f(\tau_q,r_Z)+\sum_l\tilde{\kappa}_lQ_l\upsilon_lB_f(\tau_l,r_Z),
\end{eqnarray}
where $\tau_i=4m_i^2/m_h^2$, $\upsilon_l$ are the vector couplings of the $Z$ boson to the leptons and $Q_l$ represent the charge of leptons. $\tilde{\kappa}_q$ and $\tilde{\kappa}_l$ are the effective Higgs couplings to the quarks and the leptons respectively. $C_{\gamma Z}^{SM}$ and $\widetilde{C}_{\gamma Z}^{SM}$ are the SM contributions to $h\rightarrow \gamma Z $. $C_{\gamma Z}^{NP}$ and $\widetilde{C}_{\gamma Z}^{NP}$ are the NP contributions to $h\rightarrow \gamma Z $. We can find the loop functions $A_f$, $A_W^{\gamma Z}$ and $B_f$ in Refs. \cite{HVZ-sm1,HVZ-sm2,HVZ-sm3}.
\begin{figure}
\setlength{\unitlength}{1mm}
\begin{minipage}[c]{0.8\textwidth}
\centering
\includegraphics[width=5.5in]{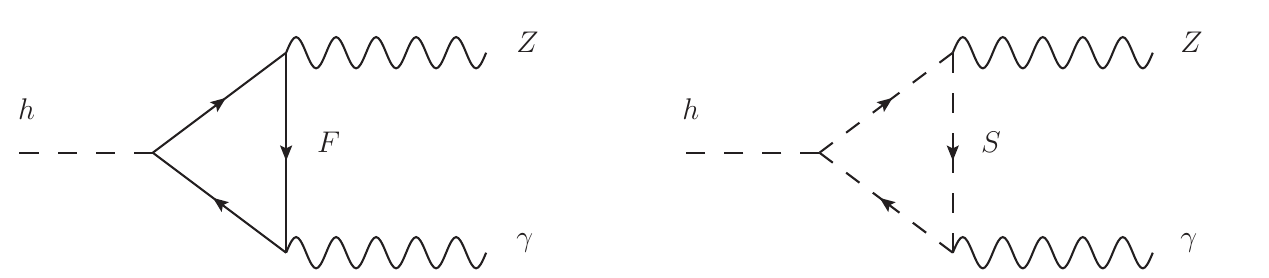}
\end{minipage}%
\caption[]{The one loop diagrams for $h\rightarrow \gamma Z$ in the $\mu\nu$SSM, with $F=\chi^{\pm}$  denoting charged fermions and $S=\tilde{u}_{i}^{+},\tilde{d}_{i}^{-},S_{\alpha}^{\pm}$ denoting squarks and charged scalars.}
\label{oneloop}
\end{figure}

In Fig. \ref{oneloop}, we show the one loop diagrams of $h\rightarrow \gamma Z$ in the $\mu\nu$SSM, where the NP contributions of $C_{\gamma Z}$ originate from the charginos, squarks and charged scalars. For the decay width of the process $h\rightarrow \gamma Z$, the QCD corrections are only about 0.1\% \cite{QCD4}, so the QCD corrections can be neglected. Although Fig.~\ref{HVZ-Feynman} and Fig. \ref{oneloop} here are similar with the MSSM, the mixing of the neutral components of the Higgs doublets with the left- and right-handed sneutrinos in the $\mu\nu$SSM can affect the mass and the coupling of the lightest Higgs boson, which can give new effect to the Higgs sector. Actually, we have discussed that the VEVs of the left-handed sneutrinos are generally small, so the mixing with the left-handed sneutrinos is not important. We can mainly consider the mixing with the right-handed sneutrinos.

In the SM, the CP-odd coupling $\widetilde{C}_{\gamma Z}^{SM}$ is 0. In the NP, $h\gamma Z$ interaction with CP-even and CP-odd parts can be written as $\bar{F}_2i(A+B\gamma_{5})F_1h$, where $A$ is the CP-even part and $B$ is the CP-odd part~\cite{HVZ-sm,HVZ-zhao}. Considered the interaction $\bar{F}_2i(C^{L}P_{L}+C^{R}P_{R})F_1h$ with $P_{L}={{1-\gamma_{5}}\over{2}}$ and $P_{R}={{1+\gamma_{5}}\over{2}}$, the CP-even part can be $A={1\over2}(C^{L}+C^{R})$ and the CP-odd part is $B={1\over2}(C^{L}-C^{R})$. In the $\mu\nu$SSM, $C_{hS_{\alpha}^{+}S_{\alpha}^{-}}^{L}=C_{hS_{\alpha}^{+}S_{\alpha}^{-}}^{R}$, $C_{h\tilde{f}\tilde{f}}^{L}=C_{h\tilde{f}\tilde{f}}^{R}$ and $C_{h\widetilde{\chi}^{\pm}_{i}\widetilde{\chi}^{\mp}_{i}}^{L}=C_{h\widetilde{\chi}^{\pm}_{i}\widetilde{\chi}^{\mp}_{i}}^{R}$, only $C_{Z\widetilde{\chi}^{\pm}_{i}\widetilde{\chi}^{\mp}_{i}}^{L}\neq C_{Z\widetilde{\chi}^{\pm}_{i}\widetilde{\chi}^{\mp}_{i}}^{R}$ but $C_{Z\widetilde{\chi}^{\pm}_{i}\widetilde{\chi}^{\mp}_{i}}^{L}+C_{Z\widetilde{\chi}^{\pm}_{i}\widetilde{\chi}^{\mp}_{i}}^{R}\gg C_{Z\widetilde{\chi}^{\pm}_{i}\widetilde{\chi}^{\mp}_{i}}^{L}-C_{Z\widetilde{\chi}^{\pm}_{i}\widetilde{\chi}^{\mp}_{i}}^{R}$. So the CP-odd coupling $\widetilde{C}_{\gamma Z}^{NP}$ in  the $\mu\nu$SSM can be neglected approximatively. Here, we can give the expression of CP-even coupling $C_{\gamma Z}^{NP}$ in the $\mu\nu$SSM
\begin{eqnarray}
&&C_{\gamma Z}^{NP}=\frac{c_{W}}{2}[(2c_{W}^{2}-1)g_{hS_{\alpha}^{+}S_{\alpha}^{-}} \frac{m_{Z}^{2}}{m_{S_{\alpha}^{\pm}}^{2}}
F_{0}(x_{S_{\alpha}^{\pm}},\lambda_{S_{\alpha}^{\pm}})\nonumber\\
&&\qquad\quad+\sum_{\tilde{f_{i}}=\tilde{u}_{i}^{+},\tilde{d}_{i}^{-}}N_{c}Q_{\tilde{f_{i}}}\hat{\upsilon}_{\tilde{f_{i}}}g_{h\tilde{f_{i}}\tilde{f_{i}}}
\frac{m_{Z}^{2}}{m_{\tilde{f_{i}}}^{2}}F_{0}(x_{\tilde{f_{i}}},\lambda_{\tilde{f_{i}}})\nonumber\\
&&\qquad\quad+\sum_{m,n=L,R}g_{h\widetilde{\chi}^{\pm}_{i}\widetilde{\chi}^{\mp}_{i}}^{m}g_{Z\widetilde{\chi}^{\pm}_{i}\widetilde{\chi}^{\mp}_{i}}^{n}\frac{2m_{W}}{m_{\widetilde{\chi}^{\pm}_{i}}}F_{1/2}(x_{\widetilde{\chi}^{\pm}_{i}},\lambda_{i})],
\end{eqnarray}
where $x_{i}={4m_{i}^2/ m_{h}^2}$, $\lambda_{i}={4m_{i}^2/ m_{Z}^2}$, $\hat{v}_{\tilde{f}_1}=(T_3^{f}\cos^2 \theta_f-Q_f s_W^2)/ c_{W}$, $\hat{v}_{\tilde{f}_2}=(T_3^{f}\sin^2 \theta_f-Q_f s_W^2)/ c_{W}$, $\hat{v}_{\tilde{f}_1}$ and $\hat{v}_{\tilde{f}_2}$ represent up and down-quark sectors, $T_3^{f}$ is the weak isospin of a fermion $f$, $\theta_f$ is the mixing angle of sfermions $\tilde{f}_{1,2}$, and $c_W=\cos\theta_W$. The form factors $F_0$, $F_{1/2}$, and the concrete expressions of couplings $g_{hS_{\alpha}^{+}S_{\alpha}^{-}}=-{{\upsilon}\over{2m_{Z}^2}}C_{hS_{\alpha}^{+}S_{\alpha}^{-}}^{L,R}$, $g_{h\tilde{f}\tilde{f}}=C_{h\tilde{f}\tilde{f}}^{L,R}$, $g_{h\widetilde{\chi}^{\pm}_{i}\widetilde{\chi}^{\mp}_{i}}^{L,R}=-{{1}\over{e}}C_{h\widetilde{\chi}^{\pm}_{i}\widetilde{\chi}^{\mp}_{i}}^{L,R}$, $g_{Z\widetilde{\chi}^{\pm}_{i}\widetilde{\chi}^{\mp}_{i}}^{L,R}=-{{1}\over{e}}C_{Z\widetilde{\chi}^{\pm}_{i}\widetilde{\chi}^{\mp}_{i}}^{L,R}$, can be seen in Refs.~\cite{HZrr,muon,MASS,HZr,Zhang1}.

Compared to the indirect contributions, the direct contributions to the decay amplitudes can be calculated in a power series in $(\Lambda_{QCD}/m_{h})^2$ or $(m_{q}/m_{h})^2$ \cite{HVZ-sm,HVZ-zhao}. Here, the $\Lambda_{QCD}$ is a hadronic scale where  the authors in Ref. \cite{QCD5} provide $\Lambda_{QCD}/m_Z\sim0.01$, and $m_{q}$ are the effective masses of the constituent quarks of a given meson. The asymptotic function $\phi_{M}^{\perp}(x)=6x(1-x)$ \cite{HVZ-zhao,phi-function1,phi-function2,phi-function3} is needed, then the direct contributions are as follow,
\begin{eqnarray}
F_{\bot {direct}}^{MZ}=\sum\limits_{q}f_{M}^{q\bot}\upsilon_{q}\kappa_{q}{{3m_q}\over{2m_M}}{{1-r_{Z}^2
+2r_{Z}\ln{r_{Z}}}\over{(1-r_{Z})^2}},\\
\tilde{F}_{\bot {direct}}^{MZ}=\sum\limits_{q}f_{M}^{q\bot}\upsilon_{q}\tilde{\kappa}_{q}
{{3m_q}\over{2m_M}}{{1-r_{Z}^2+2r_{Z}\ln{r_{Z}}}\over{(1-r_{Z})^2}}.
\end{eqnarray}
Here, $f_{M}^{q\bot}$ are the flavor-specific transverse decay constants of the meson \cite{HVZ-sm3}. In our calculations, we found that the direct contribution is small compared with the indirect contributions. This type of direct contributions are strongly suppressed, and there is no NP windows. The so-called direct contributions to the decay amplitudes involve the Yukawa couplings of the valence quarks in the meson $M$ and are typically subdominant~\cite{HVZ-sm}. So, searching the NP on the Yukawa coupling of the light quarks is not suitable, and the indirect contributions are more important than the direct contributions in our study.

Normalized to the SM expectation, the signal strengths for the Higgs decay channels are quantified by the ratios \cite{HZrr,signal}
\begin{eqnarray}
&&\mu_{MZ}^{\rm{ggF}}={{\sigma_{\text{NP}}(\text{ggF})\text{Br}_{\rm{NP}}(h\rightarrow MZ)}\over{\sigma_{\rm{SM}}({\rm{ggF}}){{\rm{Br}}_{\rm{SM}}}(h\rightarrow MZ)}},
\label{eq}
\\
&&\mu_{\gamma\gamma}^{\rm{ggF}}={{\sigma_{\rm{NP}}({\rm{ggF}}){\rm{Br}}_{\rm{NP}}(h\rightarrow \gamma\gamma)}\over{\sigma_{\rm{SM}}({\rm{ggF}}){\rm{Br}}_{\rm{SM}}(h\rightarrow \gamma\gamma)}},
\label{eq1}
\end{eqnarray}
where the ggF stands for gluon-gluon fusion and $M=\rho, \omega, \phi, J/\psi, \Upsilon$. We evaluate the Higgs production cross sections
\begin{eqnarray}
{{\sigma_{\rm{NP}}(\rm{ggF})}\over{\sigma_{\rm{SM}}(\rm{ggF})}}
&&\approx{{\Gamma_{\rm{NP}}(h\rightarrow gg)}\over\Gamma_{\rm{SM}}(h\rightarrow gg)}={{\Gamma_{\rm{NP}}^{h}}\over{\Gamma_{\rm{SM}}^{h}}}{{\Gamma_{\rm{NP}}(h\rightarrow gg)/\Gamma_{\rm{NP}}^{h}}\over{\Gamma_{\rm{SM}}(h\rightarrow gg)/\Gamma_{\rm{SM}}^{h}}},
\label{eq3}\\
&&={{\Gamma_{\rm{NP}}^{h}}\over{\Gamma_{\rm{SM}}^{h}}}{{{\rm{Br}_{\rm{NP}}}(h\rightarrow gg)}\over{{\rm{Br}_{\rm{SM}}}(h\rightarrow gg)}},
\label{eq2}
\end{eqnarray}
where $\Gamma_{\rm{NP}}^{h}$ and $\Gamma_{\rm{SM}}^{h}$ denote the NP and SM Higgs total decay width. Through Eqs.(\ref{eq}-\ref{eq1}) and Eq.(\ref{eq3}), we quantify the signal strengths for $h\rightarrow {MZ}$ and $h\rightarrow \gamma\gamma$
\begin{eqnarray}
\mu_{{MZ}}^{\rm{ggF}}&&\approx{{\Gamma_{\rm{NP}}(h\rightarrow gg)}\over{\Gamma_{\rm{SM}}}(h\rightarrow gg)}{{\Gamma_{\rm{NP}}(h\rightarrow MZ)/\Gamma_{\rm{NP}}^{h}}\over{\Gamma_{\rm{SM}}(h\rightarrow MZ)/\Gamma_{\rm{SM}}^{h}}},
\\
&&={{\Gamma_{\rm{SM}}^{h}}\over{\Gamma_{\rm{NP}}^{h}}}{{\Gamma_{\rm{NP}}(h\rightarrow gg)}\over{\Gamma_{\rm{SM}}(h\rightarrow gg)}}{{\Gamma_{\rm{NP}}(h\rightarrow MZ)}\over{\Gamma_{\rm{SM}}(h\rightarrow MZ)}} ,
\label{muss}\\
\mu_{\gamma\gamma}^{\rm{ggF}}&&\approx{{\Gamma_{\rm{NP}}(h\rightarrow gg)}\over{\Gamma_{\rm{SM}}}(h\rightarrow gg)}{{\Gamma_{\rm{NP}}(h\rightarrow \gamma\gamma)/\Gamma_{\rm{NP}}^{h}}\over{\Gamma_{\rm{SM}}(h\rightarrow \gamma\gamma)/\Gamma_{\rm{SM}}^{h}}},
\\
&&={{\Gamma_{\rm{SM}}^{h}}\over{\Gamma_{\rm{NP}}^{h}}}{{\Gamma_{\rm{NP}}(h\rightarrow gg)}\over{\Gamma_{\rm{SM}}(h\rightarrow gg)}}{{\Gamma_{\rm{NP}}(h\rightarrow \gamma\gamma)}\over{\Gamma_{\rm{SM}}(h\rightarrow \gamma\gamma)}}.
\end{eqnarray}
We note that the signal strength is used as a guideline in the ATLAS and CMS experiments \cite{signal}.

\section{Numerical Results\label{sec4}}
In this section, we will discuss the numerical results. Firstly, we make the minimal flavor violation (MFV) \cite{MFV} assumptions for some parameters, which assume
\begin{eqnarray}
&&\hspace{-0.9cm}{\kappa _{ijk}} = \kappa {\delta _{ij}}{\delta _{jk}}, \quad
{({A_\kappa }\kappa )_{ijk}} =
{A_\kappa }\kappa {\delta _{ij}}{\delta _{jk}}, \quad
\lambda _i = \lambda , \nonumber\\
&&\hspace{-0.9cm}
{({A_\lambda }\lambda )}_i = {A_\lambda }\lambda,\quad
{Y_{{e_{ij}}}} = {Y_{{e_i}}}{\delta _{ij}},\quad
{({A_e}{Y_e})_{ij}} = {A_{e}}{Y_{{e_i}}}{\delta _{ij}},\nonumber\\
&&\hspace{-0.9cm}
{Y_{{\nu _{ij}}}} = {Y_{{\nu _i}}}{\delta _{ij}},\quad
(A_\nu Y_\nu)_{ij}={a_{{\nu_i}}}{\delta _{ij}},\quad
m_{\tilde \nu_{ij}^c}^2 = m_{\tilde \nu_{i}^c}^2{\delta _{ij}}, \nonumber\\
&&\hspace{-0.9cm}m_{\tilde Q_{ij}}^2 = m_{{{\tilde Q_i}}}^2{\delta _{ij}}, \quad
m_{\tilde u_{ij}^c}^2 = m_{{{\tilde u_i}^c}}^2{\delta _{ij}}, \quad
m_{\tilde d_{ij}^c}^2 = m_{{{\tilde d_i}^c}}^2{\delta _{ij}}, \nonumber\\
&&\hspace{-0.9cm}m_{{{\tilde L}_{ij}}}^2 = m_{{\tilde L}}^2{\delta _{ij}}, \quad
m_{\tilde e_{ij}^c}^2 = m_{{{\tilde e}^c}}^2{\delta _{ij}}, \quad
\upsilon_{\nu_i^c}=\upsilon_{\nu^c},
\label{MFV}
\end{eqnarray}
where $i,\;j,\;k =1,\;2,\;3 $. $m_{\tilde \nu_i^c}^2$ can be constrained by the minimization conditions of the neutral scalar potential seen in Ref.~\cite{MASS}. To agree with experimental observations on quark mixing, one can have\cite{Zhang1,muon, HZrr,MASS}
\begin{eqnarray}
&&\hspace{-0.75cm}\;\,{Y_{{u _{ij}}}} = {Y_{{u _i}}}{V_{L_{ij}}^u},\quad
 (A_u Y_u)_{ij}={A_{u_i}}{Y_{{u_{ij}}}},\nonumber\\
&&\hspace{-0.75cm}\;\,{Y_{{d_{ij}}}} = {Y_{{d_i}}}{V_{L_{ij}}^d},\quad
(A_d Y_d)_{ij}={A_{d}}{Y_{{d_{ij}}}},
\end{eqnarray}
and $V=V_L^u V_L^{d\dag}$ denotes the CKM matrix \cite{PDG1,CKM1,CKM2}.
\begin{eqnarray}
{Y_{{u_i}}} = \frac{{{m_{{u_i}}}}}{{{\upsilon_u}}},\qquad {Y_{{d_i}}} = \frac{{{m_{{d_i}}}}}{{{\upsilon_d}}},\qquad {Y_{{e_i}}} = \frac{{{m_{{l_i}}}}}{{{\upsilon_d}}},
\end{eqnarray}
where the $m_{u_{i}},m_{d_{i}}$ and $m_{l_{i}}$ stand for the up-quark, down-quark and charged lepton masses, and we can found the value of the masses from Particle Data Group (PDG)~\cite{PDG1}. In our previous work~\cite{neu-mass6}, the Yukawa couplings $Y_{\nu_i} \sim \mathcal{O}(10^{-7,-6})$ and left-handed sneutrino VEVs $\upsilon_{\nu_i} \sim \mathcal{O}(10^{-4}\,{\rm{GeV}})$ are determined by the TeV seesaw mechanism. In the following, we could reasonably neglect the small terms including $Y_{\nu_i}$ or $\upsilon_{\nu_i}$ in the Higgs sector.

Through analysis of the parameter space of the $\mu\nu$SSM in Ref.~\cite{mnSSM1,parameter-space}, we set the parameter values to be $A_{\kappa}=-300$ GeV and $A_{u_{1,2}}=A_{d}=A_{e}=1$~TeV for simplification, which don't affect the numerical results in the following. In terms of experimental observations, the first and second generations of squarks are strongly constrained by direct searches at the LHC. In R-Parity violating model, the masses of the first and second generations of squarks at least greater than 1.6 TeV in 95\% confidence level \cite{PDG1}. Therefore, we take heavier squark mass $m_{\tilde{Q}_{1,2,3}}=m_{\tilde{u}_{1,2}^{c}}=m_{\tilde{d}_{1,2,3}^{c}}=3$~TeV. The sleptons do not have strong experimental constraints like squarks, we take $m_{\tilde{L}}=m_{\tilde{e}^{c}}=1$ TeV. We will choose the gauginos' Majorana masses $M_{1}=M_2$ for simplicity. For the R-parity violating scenario on a set of simplified SUSY models, the gluino mass $m_{\tilde{g}}\approx M_{3}$, is larger than about 2.26 TeV  \cite{PDG1,ATLAS-mg}. To be on the safe side, we take $M_{3}=2.5$ TeV in the R-parity violating $\mu\nu$SSM for safety.

\subsection{Mass and couplings of the 125 GeV Higgs boson}

In the large $m_A$ limit ($A$ is the MSSM-like pseudo-scalar), we give an approximate expression for the lightest Higgs boson mass in the $\mu\nu$SSM,~\cite{MASS}
\begin{eqnarray}
m_h^2 \approx  \xi_h ( m_{h_{\rm{MSSM}}}^2 + {6\lambda^2 s_{W}^2c_{W}^2\over e^2}m_{Z}^2 \sin^2{2\beta}),
\label{mh-app1}
\end{eqnarray}
where
\begin{eqnarray}
&&m_{h_{\rm{MSSM}}}^2 = m_{Z}^2 \cos{2\beta}^2 + \Delta m_{h_{\rm{MSSM}}}^2,\\
\label{MH1}
&&\xi_h = 1- \frac{(A_{X_1}^2)^2}{m_{R_1}^2 ( m_{h_{\rm{MSSM}}}^2 + {6\lambda^2 s_{W}^2c_{W}^2\over e^2}m_{Z}^2 \sin^2{2\beta})}\,.
\label{MH2}
\end{eqnarray}
Here $m_{h_{\rm{MSSM}}}$ is the mass of lightest Higgs boson in the MSSM, and $\Delta m_{h_{\rm{MSSM}}}^2$ is the two-loop radiative corrections, $e$ is the electromagnetic coupling constant. $A_{X_1}^2$ comes from the mixing of the neutral components of the Higgs doublets with the right-handed sneutrinos, and $m_{R_1}^2$ is the mass squared of the right-handed sneutrino, whose concrete expressions are given by
\begin{eqnarray}
&&A_{X_1}^2\simeq\sqrt{3}\lambda\upsilon\sin{2\beta}\left[2\upsilon_{\nu^{c}}\left({{3\lambda}\over{\sin{2\beta}}}-\kappa\right)-A_{\lambda}+{1\over2}(\Delta_{1R}+\Delta_{2R}) \right],
\label{AX1}\\
&&m_{R_1}^2= (A_\kappa+4\kappa\upsilon_{\nu^c})\kappa\upsilon_{\nu^c} +A_\lambda \lambda \upsilon_d \upsilon_u/\upsilon_{\nu^c} + \lambda^2 (2 \upsilon^2+3\Delta_{RR}),
\end{eqnarray}
where $\Delta_{1R}$, $\Delta_{2R}$ and $\Delta_{RR}$ are the radiative corrections~\cite{MASS}.

Compared with the MSSM, the lightest Higgs boson mass in the $\mu\nu$SSM gets an additional term ${6\lambda^2 s_{W}^2c_{W}^2\over e^2}m_{Z}^2 \sin^2{2\beta}$ \cite{tree-level,tree-level1}. For $\lambda$ $\lesssim\mathcal{O}$$(0.1)$ with $\tan\beta=5$, the additional tree-level contribution to the lightest Higgs boson mass \cite{naturalness1,tree-level2} is weak with $\mathcal{O}$$(\lesssim 1{\rm{GeV}})$, which is different from the large $\lambda$ case. In addition, we consider the mixing of the neutral components of the Higgs doublets with the right-handed sneutrinos. The mixing term $A_{X_1}$ in Eq. (\ref{AX1}) can be affected by the parameters $\lambda, \tan\beta, \upsilon_{\nu^{c}}, \kappa, A_{\lambda}$. At the same time, one can choose suitable value of the parameter in Eq.~(\ref{AX1}) to accommodate the 125 GeV lightest Higgs boson, through affecting the mixing of the neutral components of the Higgs doublets with the right-handed sneutrinos.

\begin{figure}
\setlength{\unitlength}{1mm}
\begin{minipage}[c]{0.70\textwidth}
\centering
\includegraphics[width=4.5in]{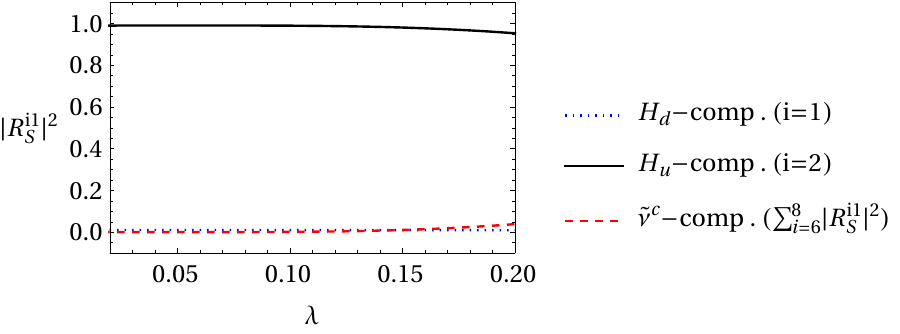}
\end{minipage}%
\caption[]{$H_{d}$-component, $H_{u}$-component and $\tilde{\nu}^{c}$-component of the lightest Higgs boson versus the parameter $\lambda$, which can be defined by $|R_{S}^{i1}|^2$ with $i=1,2$ and $\Sigma^8_{i=6}|R_{S}^{i1}|^2$ respectively. The unitary matrix $R_{S}$ is used to diagonalize the CP-even neutral scalar mass matrix, which can be found in Ref. \cite{Zhang1}.}
\label{huhd}
\end{figure}

In special parameter space, the lightest Higgs boson in the $\mu\nu$SSM can have more than 40\% $\tilde{\nu}^{c}$-component~\cite{Hu-vc},  which may be not an SM-like Higgs boson. In this work, we plot $H_{d}$-component, $H_{u}$-component and $\tilde{\nu}^{c}$-component of the lightest Higgs boson varying with the parameter $\lambda$ in Fig. \ref{huhd}, taking $M_2=500$~GeV, $\kappa=0.3$, $\upsilon_{\nu^c}=1$~TeV, $A_t=3$~TeV, $m_{{\tilde u}_3^c}=3$~TeV, $\tan\beta=10$ and $A_{\lambda}=1$~TeV. We can see that the lightest Higgs boson can be a mixed state with about 98\% $H_{u}$-component and 2\% $\tilde{\nu}_{c}$-component, when $\lambda=0.2$. As $\lambda<0.15$, the lightest Higgs boson can have over 99\% $H_{u}$-component, which is an SM-like Higgs boson. Note that $H_{u}$-component of the lightest Higgs boson is around 99\% in the following numerical results.

The parameters $A_{u_{3}}=A_t$, $m_{\tilde{u}_{3}^{c}}$ and $\tan\beta$ together with $\lambda$,  $\upsilon_{\nu^{c}}$, ${\kappa}$ and ${A_{\lambda}}$ can deeply affect the lightest Higgs boson mass and couplings. The contributions of charginos account for a large proportion in NP. We focus on the impact of the new parameters $\lambda$ and $\upsilon_{\nu^{c}}$ in the $\mu\nu$SSM, which can change chargino mass and the coupling of Higgs boson and chargino.

In the MSSM, the Higgs coupling to chargino pair can be written as \cite{MSSM4}
\begin{eqnarray}
C_{h\widetilde{\chi}_{i}^{-}\widetilde{\chi}_{j}^{+}}^{L, MSSM}=-{{e}\over{\sqrt{2}s_{W}}}(Z_{R}^{11}Z_{M+}^{1i}Z_{M-}^{2j}+
Z_{R}^{21}Z_{M+}^{2i}Z_{M-}^{1j}),
\label{coupling}
\end{eqnarray}
where $Z_{R}$, $Z_{M+}$ and $Z_{M-}$ are the unitary  matrices in the MSSM.
The coupling of Higgs boson and chargino in the $\mu\nu$SSM is \cite{HZrr}
\begin{eqnarray}
C_{h\widetilde{\chi}_{i}^{-}\widetilde{\chi}_{j}^{+}}^{L, \mu\nu SSM}=
&&-{{e}\over{\sqrt{2}s_{W}}}[R_{S}^{11}Z_{+}^{1i}Z_{-}^{2j}
+R_{S}^{21}Z_{+}^{2i}Z_{-}^{1j}+R_{S}^{(2+k)1}Z_{+}^{1i}Z_{-}^{(2+k)j}]
\nonumber\\
&&-{{Y_{e_{kl}}}\over{\sqrt{2}}}[R_{S}^{11}Z_{+}^{(2+k)i}Z_{-}^{(2+l)j}
-R_{S}^{(2+k)1}Z_{+}^{(2+l)i}Z_{-}^{1j}] \nonumber\\
&&-{{Y_{\nu_{kl}}}\over{\sqrt{2}}}R_{S}^{(5+k)1}Z_{+}^{2i}Z_{-}^{(2+l)j}
-{{\lambda_{k}}\over{\sqrt{2}}}R_{S}^{(5+k)1}Z_{+}^{2i}Z_{-}^{2j},
\label{coupling1}
\end{eqnarray}
where the unitary matrices $Z_{+}, Z_{-}$ are used to diagonalize the charged fermion mass matrix, which can be found in Ref. \cite{Zhang1}. Compared to the coupling $C_{h\widetilde{\chi}_{i}^{-}\widetilde{\chi}_{j}^{+}}^{L,MSSM}$ in the MSSM, the coupling $C_{h\widetilde{\chi}_{i}^{-}\widetilde{\chi}_{j}^{+}}^{L,\mu\nu SSM}$ in the $\mu\nu$SSM has five extra terms, which can give new contributions to the coupling.

\begin{figure}
\setlength{\unitlength}{1mm}
\begin{minipage}[c]{0.50\textwidth}
\centering
\includegraphics[width=3.5in]{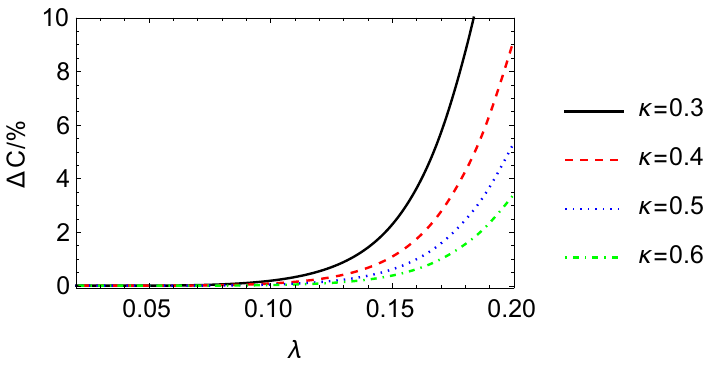}
\end{minipage}%
\caption[]{$\Delta C$ varying with $\lambda$ for different $\kappa$.}
\label{C}
\end{figure}

The last term of Eq. (\ref{coupling1}) can be directly affected by $\lambda_k$, we will focus on analyzing how much contribution will be brought by the last term in Eq.~(\ref{coupling1}) compared to the MSSM, defining $\Delta C={{{{\lambda_{k}}\over{\sqrt{2}}}R_{S}^{(5+k)1}Z_{+}^{2i}Z_{-}^{2j}}/{C_{h\widetilde{\chi}_{i}^{-}\widetilde{\chi}_{j}^{+}}^{L, MSSM}}}$. We plot $\Delta C$ versus $\lambda_k\equiv\lambda$ for different $\kappa$ in Fig.~\ref{C}, when  $M_2=500$ GeV, $\upsilon_{\nu^c}=1$ TeV, $A_t=3$ TeV, $m_{{\tilde u}_3^c}=3$ TeV, $\tan\beta=10$ and $A_{\lambda}=1$ TeV. In Fig. \ref{C}, one can know that $\Delta C$ increases with increasing of $\lambda$. When $\lambda=0.2$, $\Delta C\approx 3.5\%$ for $\kappa=0.6$, $\Delta C\approx 5\%$ for $\kappa=0.5$, $\Delta C\approx 9\%$ for $\kappa=0.4$, $\Delta C> 10\%$ for $\kappa=0.3$, respectively. Even the term ${{{\lambda_{k}}\over{\sqrt{2}}}R_{S}^{(5+k)1}Z_{+}^{2i}Z_{-}^{2j}}$ is $\propto$ $\lambda_k$, the contribution is still suppressed by $R_{S}^{(5+k)1}$, especially when $\lambda_k<0.1$. Furthermore, due that $Y_{e_{kl}}$ and $Y_{\nu_{kl}}$ are tiny, these terms ${{Y_{e_{kl}}}\over{\sqrt{2}}}[R_{S}^{11}Z_{+}^{(2+k)i}Z_{-}^{(2+l)j}-R_{S}^{(2+k)1}Z_{+}^{(2+l)i}Z_{-}^{1j}]$ and ${{Y_{\nu_{kl}}}\over{\sqrt{2}}}R_{S}^{(5+k)1}Z_{+}^{2i}Z_{-}^{(2+l)j}$ are almost negligible.

The soft mass $m_{H_{u}}^{2}$ is negative in this work. For example, the soft mass is about $-0.5< m_{H_{u}}^{2} (\textrm{TeV}^2)< -0.1$, on parameter space for $\tan\beta=5, \kappa=0.5, A_{\lambda}=1$ TeV, $\upsilon_{\nu^c}=2$ TeV, and $0.1<\lambda<0.2$. It is also useful to mention that the parameter $B_{\mu}$ is written as $B_{\mu}=3A_{\lambda}\lambda \upsilon_{\nu^c}=\mu A_{\lambda}$ in the $\mu\nu$SSM.

So in our next analysis, we will consider the following parameters
\begin{eqnarray}
\tan \beta ,\; m_{{\tilde u}_3^c},\; {A_{t}}, \;  \lambda, \; {\upsilon_{\nu^c}}, \; {M_{2}}, \; {\kappa}, \; {A_{\lambda}}.
\end{eqnarray}

\begin{figure}
\setlength{\unitlength}{1mm}
\begin{minipage}[c]{0.50\textwidth}
\centering
\includegraphics[width=3.0in]{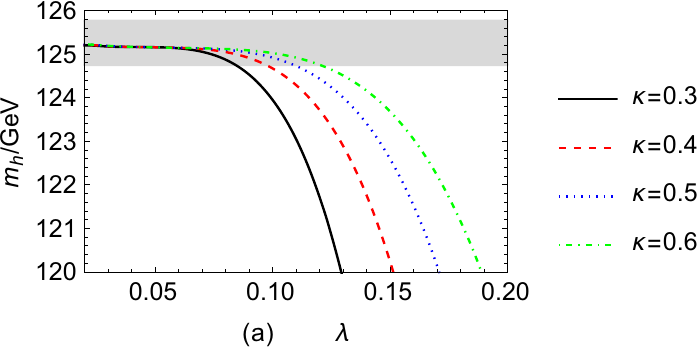}
\end{minipage}%
\begin{minipage}[c]{0.50\textwidth}
\centering
\includegraphics[width=3.0in]{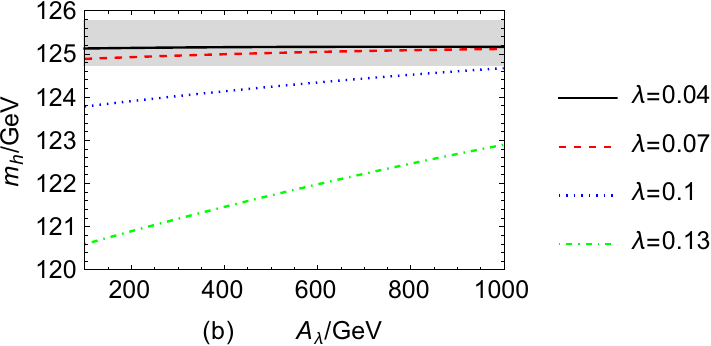}
\end{minipage}%
\caption[]{The lightest Higgs boson mass $m_h$ varying with $\lambda$ (a) and $A_{\lambda}$/GeV (b), where the grey area denotes the experimental value of $m_h$ at $3\sigma$ \cite{PDG1}. We set parameters $A_{\lambda}=1$ TeV in (a) and $\kappa=0.6$ in (b).}
\label{mh}
\end{figure}

In order to further analyze the influence of relevant parameters on the lightest Higgs boson decays, we first investigate the influence of relevant parameters on the mass and coupling of the lightest Higgs boson in the model.
Taking $M_2=500$ GeV, $\upsilon_{\nu^c}=1$ TeV, $A_t=3$ TeV, $m_{{\tilde u}_3^c}=3$ TeV, $\tan\beta=10$, we plot the lightest Higgs boson mass  $m_h$ of the $\mu\nu$SSM in Fig. \ref{mh}, where the grey area denotes the experimental value of $m_h$ at $3\sigma$ where the experimental value is that  $m_h=125.25\pm0.17$ \cite{PDG1}. In Fig. \ref{mh} (a), the lightest Higgs boson mass $m_h$ is varying with the parameter $\lambda$ in different $\kappa$, with $A_{\lambda}=1$ TeV. The numerical results show that $m_h$ in the $\mu\nu$SSM drops down very quickly, with increasing of $\lambda$. Because the mixing term $A_{X_1}$ in Eq. (\ref{AX1}) is very sensitive to the parameter $\lambda$. For $\kappa=0.3$, when the parameter $\lambda$ is larger than 0.09, the lightest Higgs boson mass $m_h$ starts under the lower bound of the grey area. If $\kappa$=0.6, $m_h$ is below the lower bound of the grey area when $\lambda>0.12$. Through Eq. (\ref{AX1}), there is a minus sign in front of the parameter $\kappa$, which reduce the mixing term on the lightest Higgs boson mass for a larger $\kappa$. But, due to constrained by Landau pole condition \cite{Landau}, here we choose the parameter $\kappa\leq$0.6 \cite{k}.

In Fig. \ref{mh} (b), we plot the lightest Higgs boson mass $m_h$ versus $A_{\lambda}$ in different $\lambda$, with $\kappa=0.6$. We can see that $m_h$ increases as $A_{\lambda}$ increases, and this behavior becomes more pronounced when $\lambda$ is greater than 0.1. For $\lambda=0.04$ or $\lambda=0.07$, $m_h$ will be gentle with increasing of $A_{\lambda}$. Because a relatively large $A_{\lambda}$ and a small $\lambda$ reduce the effect of the mixing term $A_{X_1}$ on the lightest Higgs boson mass.
If one use the relation
\begin{eqnarray}
A_{\lambda}=2\upsilon_{\nu^{c}}\left({{3\lambda}\over{\sin{2\beta}}}-\kappa\right)
+{1\over2}(\Delta_{1R}+\Delta_{2R}),
\end{eqnarray}
the mixing term $A_{X_1}=0$, and then the 125 GeV Higgs boson will not affected by the mixing of Higgs doublets and right-handed sneutrinos.
Through the mixing term $A_{X_1}$ in Eq. (\ref{AX1}), we can know that the lightest Higgs boson mass in the $\mu\nu$SSM is affected by the parameters $\lambda, \tan\beta, \upsilon_{\nu^{c}}, \kappa, A_{\lambda}$. In the following, we will scan these parameters to assure a 125 GeV Higgs mass, considered the radiative corrections.

\begin{figure}
\setlength{\unitlength}{1mm}
\begin{minipage}[c]{0.50\textwidth}
\centering
\includegraphics[width=2.9in]{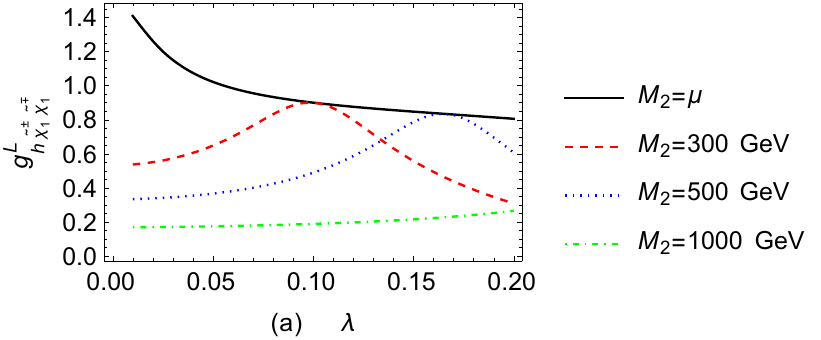}
\end{minipage}%
\begin{minipage}[c]{0.50\textwidth}
\centering
\includegraphics[width=2.9in]{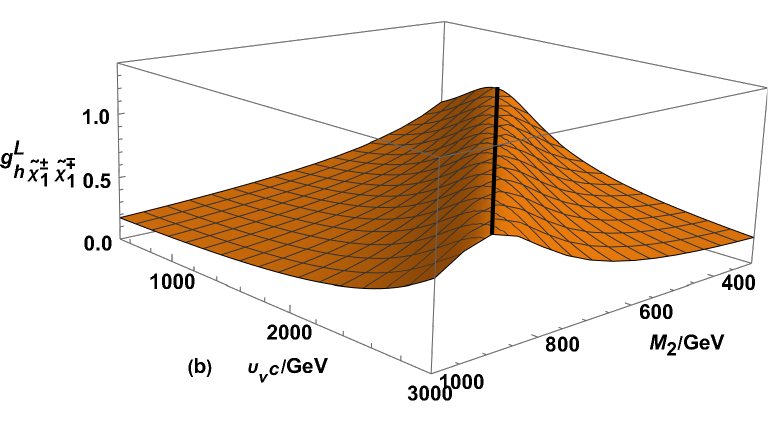}
\end{minipage}%
\caption[]{The left plot is the coupling $g^{L}_{h\widetilde{\chi}^{\pm}_{1}\widetilde{\chi}^{\mp}_{1}}$ versus the parameter $\lambda$ in different $M_{2}$. The right 3-d plot is the coupling $g^{L}_{h\widetilde{\chi}^{\pm}_{1}\widetilde{\chi}^{\mp}_{1}}$ versus the parameters $M_{2}$ and $\upsilon_{\nu^{c}}$, we set $\lambda=0.1$ and other parameters are keep same as the left plot. The black line is the $M_{2}=\mu$ case, which is also the maximum values of the coupling $g^{L}_{h\widetilde{\chi}^{\pm}_{1}\widetilde{\chi}^{\mp}_{1}}$.}
\label{gl-au}
\end{figure}

We also focus on the coupling of the lightest Higgs boson in the model, especially the coupling of the lightest Higgs boson and chargino pair $g^{L}_{h\widetilde{\chi}^{\pm}_{i}\widetilde{\chi}^{\mp}_{i}}={{1}\over{e}}C_{h\widetilde{\chi}^{\pm}_{i}\widetilde{\chi}^{\mp}_{i}}^{L}$, where $C_{h\widetilde{\chi}^{\pm}_{i}\widetilde{\chi}^{\mp}_{i}}^{L}$ can be found in Ref. \cite{Zhang1}. The parameters $\lambda, {\upsilon_{\nu^c}}$ and $M_{2}$ also affect the $R_{S}, Z_{+}, Z_{-}$ matrices.
In Fig. \ref{gl-au}, we picture $g^{L}_{h\widetilde{\chi}^{\pm}_{1}\widetilde{\chi}^{\mp}_{1}}$ versus the parameter $\lambda$ in different $M_{2}$,  with $\upsilon_{\nu^{c}}=1$ TeV, $\tan\beta=30$, $A_t=4$ TeV, $m_{{\tilde u}_3^c}= 3$ TeV, $\kappa=0.5$, $A_{\lambda}=500$ GeV. As $M_2=\mu$ with $\mu\equiv3\lambda {\upsilon_{\nu^c}}$, we can see that the coupling $g^{L}_{h\widetilde{\chi}^{\pm}_{1}\widetilde{\chi}^{\mp}_{1}}$ is about 1.4 with $\lambda=0.01$, 1.15 with $\lambda=0.03$, and 0.9 with $\lambda=0.1$, respectively.
However, fixed $M_2$, the coupling $g^{L}_{h\widetilde{\chi}^{\pm}_{1}\widetilde{\chi}^{\mp}_{1}}$  will first increase until $\lambda =M_2 / (3 {\upsilon_{\nu^c}})$ and then decrease, with increasing of $\lambda$. For $M_{2}=1000$ GeV, the maximum value of the coupling occurs when $\lambda$ is about 0.33. The numerical results show that the coupling of the lightest Higgs boson and chargino pair $g^{L}_{h\widetilde{\chi}^{\pm}_{1}\widetilde{\chi}^{\mp}_{1}}$ is more strong when $M_2=\mu$, compared that $M_2\neq \mu$. The same conclusion can be drawn from the 3-d plot on the right of Fig. \ref{gl-au}, the coupling $g^{L}_{h\widetilde{\chi}^{\pm}_{1}\widetilde{\chi}^{\mp}_{1}}$ versus the parameters $\upsilon_{\nu^{c}}, M_{2}$, When the parameter $M_{2}$ is fixed, as the $\upsilon_{\nu^{c}}$ increase the coupling becomes stronger till $M_{2}=\mu$ and then decrease.

\subsection{The decays of $h\rightarrow MZ$}
\begin{figure}
\setlength{\unitlength}{1mm}
\begin{minipage}[c]{0.50\textwidth}
\centering
\includegraphics[width=2.8in]{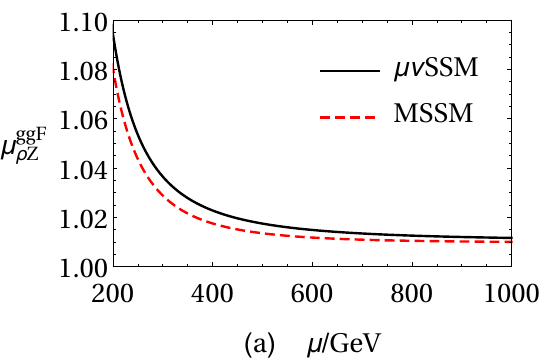}
\end{minipage}%
\begin{minipage}[c]{0.50\textwidth}
\centering
\includegraphics[width=2.8in]{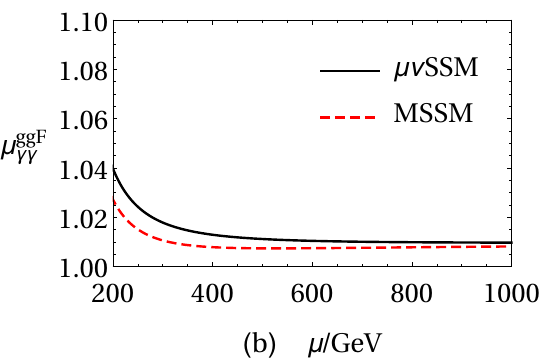}
\end{minipage}%
\caption[]{The signal strengths $\mu_{\rho Z,\gamma\gamma}^{\rm{ggF}}$ in the $\mu\nu$SSM versus the parameter $\mu$, compared to those in the MSSM. }
\label{mu}
\end{figure}
Before researching the lightest SM-like Higgs boson decays in the $\mu\nu$SSM, it's interesting to do a comparison between the signal strengths $\mu_{M Z,\gamma\gamma}^{\rm{ggF}}$ in the $\mu\nu$SSM with those in the MSSM. In Fig. \ref{mu}, we plot the signal strengths $\mu_{\rho Z,\gamma\gamma}^{\rm{ggF}}$  in the $\mu\nu$SSM versus the parameter $\mu$, compared to those in the MSSM, with $M_{2}=\mu$, $\kappa=0.5$, $A_{\lambda}=500$ GeV, $m_{\tilde{u}_{3}^{c}}=3$~TeV and $A_t=4$~TeV. In order to keep the 125 GeV Higgs boson mass in these two models that are constrained by the experimental result at 3$\sigma$, we choose  $\tan \beta =5.5$ in the MSSM, while $\tan \beta =5.4$  and $\lambda =0.05$ in the $\mu\nu$SSM. In Fig. \ref{mu}, the black solid line denotes the signal strength in the $\mu\nu$SSM, and  the red dashed line denotes the signal strength in the MSSM.

The numerical results in Fig. \ref{mu}(a) show the signal strength $\mu_{\rho Z}^{\rm{ggF}}$ in the $\mu\nu$SSM is about 1.4\% larger than that in the MSSM as $\mu=200$ GeV, and $\mu_{\rho Z}^{\rm{ggF}}$ in the $\mu\nu$SSM is about 0.5\% larger than that in the MSSM  when $\mu=400$ GeV. The right-handed  neutrino superfields introducing in the $\mu\nu$SSM lead the mixing of the neutral components of the Higgs doublets with the right-handed sneutrinos, that is different from the Higgs sector of the MSSM. The mixing can change the Higgs couplings, like the coupling of lightest Higgs boson with charginos $h\widetilde{\chi}^{\pm}_{i}\widetilde{\chi}^{\mp}_{i}$, which can give new contributions for the signal strength $\mu_{\rho Z}^{\rm{ggF}}$ in $\mu\nu$SSM. When the parameter $\mu$ grows, the coupling $C_{h\widetilde{\chi}_{i}^{-}\widetilde{\chi}_{j}^{+}}^{L,\mu\nu SSM}$ is close to $C_{h\widetilde{\chi}_{i}^{-}\widetilde{\chi}_{j}^{+}}^{L,MSSM}$. Because the charginos masses will become heavier as the parameter $\mu$ grows, the contribution from charginos to the Higgs coupling become smaller. And then, the five extra terms in Eq. (\ref{coupling1}) compared to Eq. (\ref{coupling}) can be neglect. Therefore, the signal strength $\mu_{\rho Z}^{\rm{ggF}}$ in the $\mu\nu$SSM is close to that in the MSSM, as $\mu>400$ GeV. When $\mu=1000$ GeV, the signal strength $\mu_{\rho Z}^{\rm{ggF}}$ in the $\mu\nu$SSM is only about 0.2\% larger than that in the MSSM. For the other signal strengths $\mu_{M Z}^{\rm{ggF}}$ with $M=( \omega, \phi, J/\psi, \Upsilon)$, the results in trend are similar to $\mu_{\rho Z}^{\rm{ggF}}$.

In Fig. \ref{mu}(b), the numerical results also show that the signal strength $\mu_{\gamma\gamma}^{\rm{ggF}}$ decrease, with increasing of the parameter $\mu$. Because the parameter $\mu$ also affect the mass of chargino. The lighter MSSM-like chargino can give the contributions to the signal strength $\mu_{\gamma\gamma}^{\rm{ggF}}$. In our previous work~\cite{HZrr}, the signal strength $\mu_{\gamma\gamma}^{\rm{ggF}}$ for the lighter stop and stau has been discussed in $\mu\nu$SSM. We will concretely analyse that an enhancement of the signal strengths $\mu_{M Z,\gamma\gamma}^{\rm{ggF}}$ in the $\mu\nu$SSM come from the lighter chargino mass and the Higgs couplings in the following.

In order to analyze the signal strengths $\mu_{M Z,\gamma\gamma}^{\rm{ggF}}$ in the $\mu\nu$SSM, we firstly need to choose suitable scanning range for these parameters  shown in  Table \ref{tabscan}.
Here $\upsilon_{\nu^c}\leq3$ TeV and $\lambda\leq0.2$, then $\mu$ is almost less than about 1.8 TeV. $M_2$ is taken as a free parameter to scan in the following. In this case, the parameter space makes the naturalness to be improved.
For the first generation of sfermions, naturalness requires that they only need to be below $10^4$ TeV \cite{naturalness}. For Higgs bosons, the lightest Higgs boson mass is easily affected by the stop mixing parameter  $X_t=A_t-\mu\cot\beta$, and the maximum value for $|X_t|=X_{t}^{\rm{max}}=\sqrt{6}m_{{\tilde t}_1}$, which is known as the ``maximal mixing" \cite{naturalness1}. Here $m_{{\tilde t}_1}$ denotes the lighter stop mass. Then, we choose $A_t-\mu\cot\beta\leq\sqrt{6}m_{{\tilde t}_1}$ in the scanning of Table II to constrain the value of the parameter $A_t$.

\begin{table}
\small%
\begin{tabular}{|cccc|}
\hline
Parameters  \qquad&  Min  \qquad&  Max &\\ \hline
$\tan\beta$  & 2   & 40  & \\
$m_{\tilde{u}_{3}^{c}}/\rm{TeV}$      & 1 & 5 & \\
$\upsilon_{\nu^c}/\rm{TeV}$ & 0.3 & 3 &\\
$\lambda$  &0.02 & 0.2 &\\
$A_t/\rm{TeV}$      & 1 & 5 & \\
$M_2/\rm{TeV}$      & 0.1 & 2 & \\
$\kappa$      & 0.01 & 0.6 & \\
$A_\lambda/\rm{TeV}$      & 0.1 & 1 & \\
\hline
\end{tabular}
\caption{Random scan parameters for the signal strengths $\mu_{M Z,\gamma\gamma}^{\rm{ggF}}$. We set the sleptons mass $m_{\tilde{L}}=1$ TeV.}
\label{tabscan}
\end{table}

In the scanning, there are several important experimental observed quantities which should be considered. Firstly, the results are constrained by the lightest SM-like Higgs boson mass in the $\mu\nu$SSM with $124.74\,{\rm GeV}\leq m_{{h}} \leq125.76\:{\rm GeV}$, where a $3 \sigma$ experimental error is considered \cite{PDG1}.
For the signal strengths of the lightest SM-like Higgs boson decay modes $h \rightarrow \gamma\gamma, \;WW^*, \;ZZ^*, \; b\bar b, \;\tau\bar\tau$, we adopt the averages of the results from PDG~\cite{PDG1}
\begin{eqnarray}
&&\mu_{\gamma\gamma}^{exp}=1.11_{-0.09}^{+0.10},\quad
\mu_{WW^*}^{exp}=1.19\pm0.12,\quad
\mu_{ZZ^*}^{exp}=1.06\pm0.09,\nonumber\\
&&\mu_{b\bar b}^{exp}=1.04\pm0.13,\quad
\mu_{\tau\bar\tau}^{exp}=1.15_{-0.15}^{+0.16}.
\label{hdecay}
\end{eqnarray}
In our previous work~\cite{HZrr}, the signal strengths $\mu^{ggF}_{\gamma\gamma},\mu^{ggF}_{WW^*},\mu^{ggF}_{ZZ^*}$, $\mu^{ggF}_{b\bar{b}},\mu^{ggF}_{\tau\bar{\tau}}$ contributing from the stop and stau  have been discussed in the $\mu\nu$SSM. The numerical results of the previous study show that the signal strengths in the $\mu\nu$SSM are in agreement with those in the SM, when the lighter stop mass $m_{{\tilde t}_1}> 700\;{\rm GeV}$ and the lighter stau mass $m_{{\tilde \tau}_1}> 300\;{\rm GeV}$. Constrained from the experimental limits on the stop and stau masses for the R-parity violating scenario now \cite{PDG1}, we can take $m_{\tilde{Q}_{1,2,3}}=m_{\tilde{u}_{1,2,3}^{c}}=m_{\tilde{d}_{1,2,3}^{c}}=3$~TeV and $m_{\tilde{L}}=m_{\tilde{e}^{c}}=1$ TeV for safety  considering MFV assumptions in Eq. (\ref{MFV}). As a result, the stop and stau contributions can be ignored to the signal strengths here.
In this paper, we will analyse that an enhancement of the signal strengths in the $\mu\nu$SSM come from the lighter chargino mass and the Higgs couplings in the following.

At the Large Electron-Positron collider (LEP), the chargino has been searched for in fully-hadronic, semi-leptonic and fully leptonic decay modes \cite{LEP}, which a general lower limit on the mass of chargino is 103.5 GeV. In the scanning of Table  \ref{tabscan}, we constrain the chargino mass with $m_{\tilde{\chi}^{\pm}_1}>103.5$ GeV, where ${\tilde{\chi}^{\pm}_1}$ denotes the lighter chargino leaving the charged leptons aside.

\begin{figure}
\setlength{\unitlength}{1mm}
\begin{minipage}[c]{0.5\textwidth}
\centering
\includegraphics[width=3in]{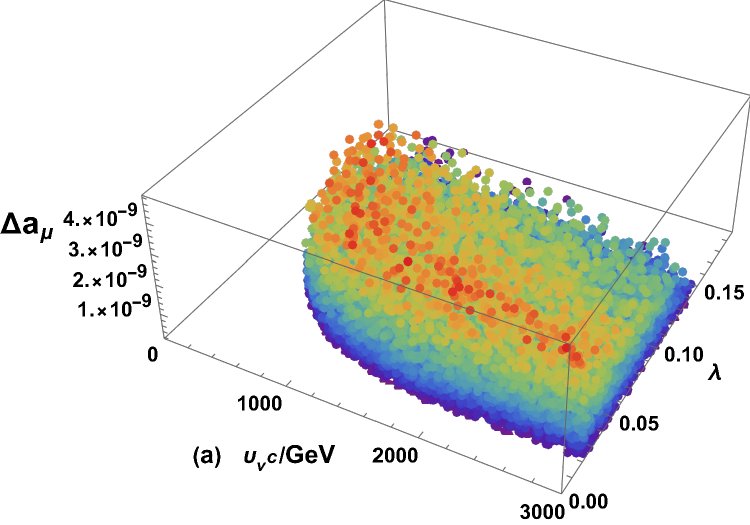}
\end{minipage}%
\begin{minipage}[c]{0.5\textwidth}
\centering
\includegraphics[width=3in]{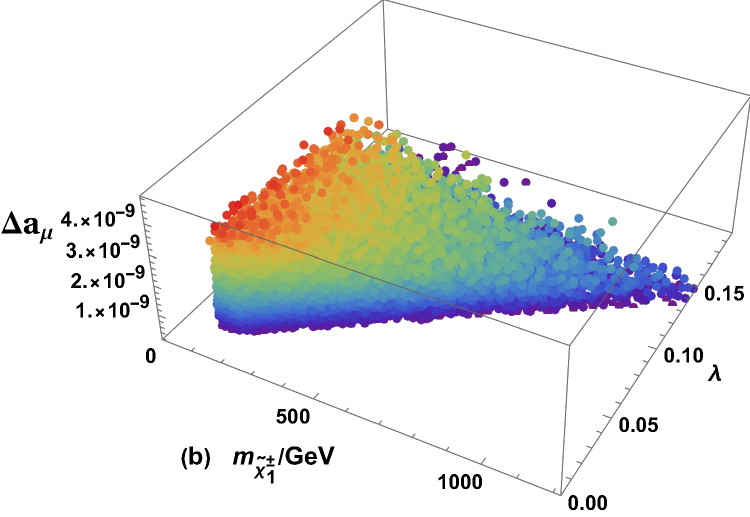}
\end{minipage}%
\caption[]{$\Delta a_{\mu}$ versus the parameters $\lambda$ and $\upsilon_{\nu^c}$ (a). The right (b) is $\Delta a_{\mu}$ versus the parameter $\lambda$ and lighter chargino mass $m_{\tilde{\chi}^{\pm}_1}$. Note that the nature of the lighter chargino ${\tilde{\chi}^{\pm}_1}$ depends on $M_{2}$ and $\mu$. The lighter chargino would be nearly pure Higgsino-like state when $M_{2}$ is much larger than $\mu$, or roughly pure Wino-like state when $M_{2}$ is much less than $\mu$, which also can be seen in Fig. \ref{HW}.}
\label{al-scan}
\end{figure}

Recently, the new experimental average for the difference between the experimental measurement and the SM theoretical prediction of the muon anomalous magnetic dipole moment (MDM) is given by \cite{g-2new}
\begin{eqnarray}
\Delta a_{\mu}=a_{\mu}^{exp}-a_{\mu}^{SM}=(25.1\pm5.9)\times10^{-10}.
\end{eqnarray}
Through our previous work \cite{muon}, the SUSY contributions of muon anomalous MDM $\Delta a_{\mu}$ in the $\mu\nu$SSM can be large in the presence of lighter sleptons, charginos or neutralinos. Through a random scan in Table II, we plot $\Delta a_{\mu}$ varying with the parameter $\lambda$ and $\upsilon_{\nu^c}/$ GeV in Fig. \ref{al-scan} (a). To show $\Delta a_{\mu}$ in gradients, we picture with the different colour  in gradients. In Fig. \ref{al-scan}, these orange and yellow dots represent large $\Delta a_{\mu}$, and blue dots represent small $\Delta a_{\mu}$. The numerical result indicates that $\Delta a_{\mu}$ can be large when $\mu=3\lambda \upsilon_{\nu^c}$ is small. Because the parameter $\mu$ can affect the chargino masses. To see more clearly, Fig. \ref{al-scan} (b) shows $\Delta a_{\mu}$ versus the parameter  $\lambda$ and the lighter chargino mass $m_{\tilde{\chi}^{\pm}_1}$. We can see that  $\Delta a_{\mu}$ decreases with increasing of $m_{\tilde{\chi}^{\pm}_1}$. When $m_{\tilde{\chi}^{\pm}_1}$ is large than about 1.2 TeV, $\Delta a_{\mu}$ is easily excluded by the new experimental average at $3\sigma$. Therefore, the following results in the scanning are also constrained by $\Delta a_{\mu}$ at $3\sigma$ level.

Be similar to the anomalous MDM of muon, the branching ratio of $\bar{B}\rightarrow X_s\gamma$  gets large  $\tan\beta$ enhancements from the down-fermion Yukawa couplings, ${Y_{{d_i}}} = {{{m_{{d_i}}}}}/{{{\upsilon_d}}} = {{{m_{{d_i}}\sqrt{\tan^2\beta+1}}}}/{{{\upsilon}}}$ and ${Y_{{e_i}}} = {{{m_{{l_i}}}}}/{{{\upsilon_d}}} = {{{m_{{l_i}}\sqrt{\tan^2\beta+1}}}}/{{{\upsilon}}}$ with $\upsilon=\sqrt{\upsilon_d^2+\upsilon_u^2}\simeq 174$ GeV.
In our previous work~\cite{bsy}, $\tan\beta$ and $A_t$ are the key parameters which affect ${\rm{Br}}(\bar{B}\rightarrow X_{s}\gamma)$ in the $\mu\nu$SSM. The current experimental value for the branching ratio of $\bar{B}\rightarrow X_{s}\gamma$ is ${\rm{Br}}(\bar{B}\rightarrow X_{s}\gamma)=(3.49\pm0.19)\times10^{-4}$ \cite{PDG1}, which will be considered $3\sigma$ experimental error in our scanning.

\begin{figure}
\setlength{\unitlength}{1mm}
\begin{minipage}[c]{0.5\textwidth}
\centering
\includegraphics[width=3in]{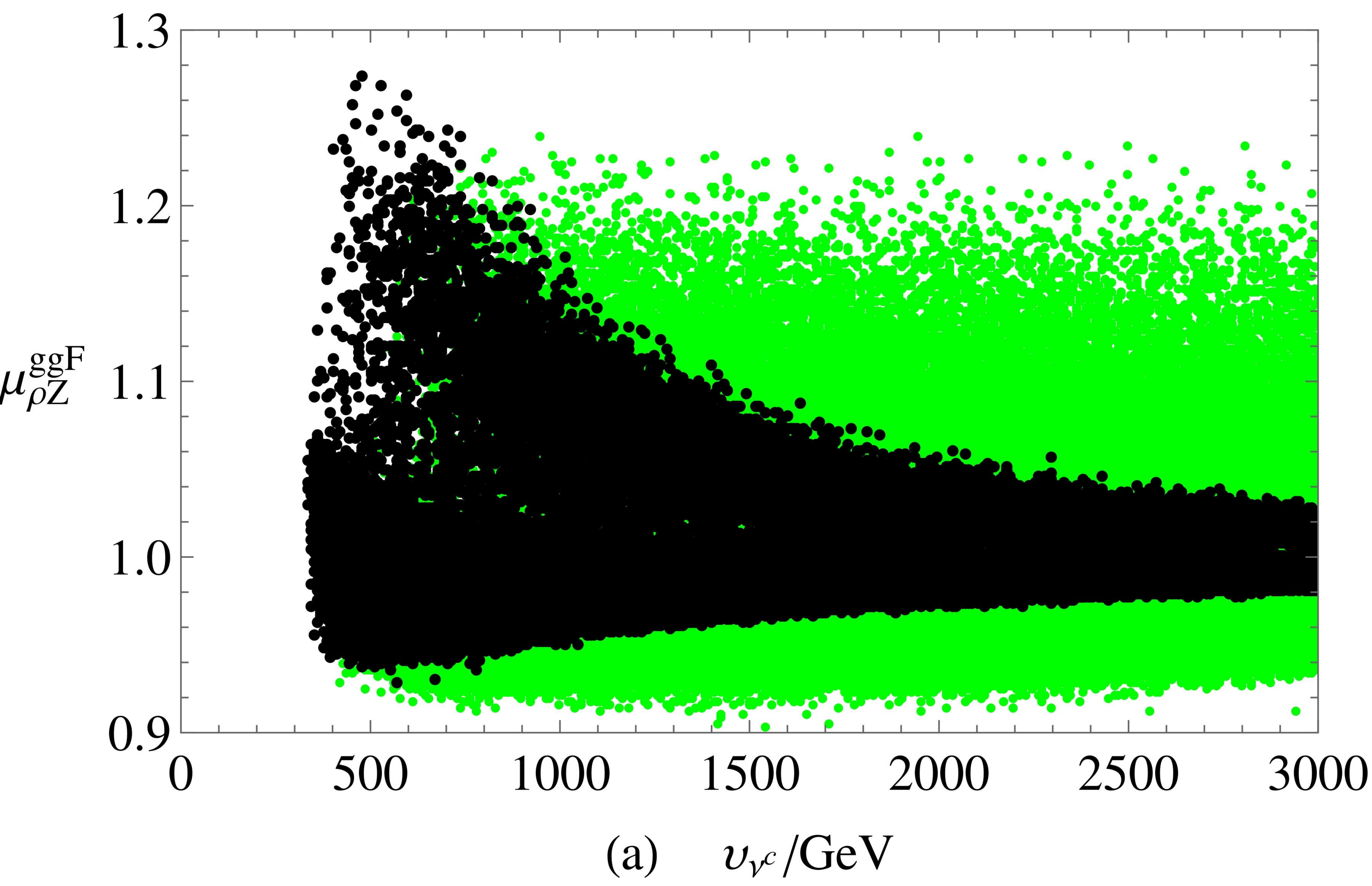}
\end{minipage}%
\begin{minipage}[c]{0.5\textwidth}
\centering
\includegraphics[width=3in]{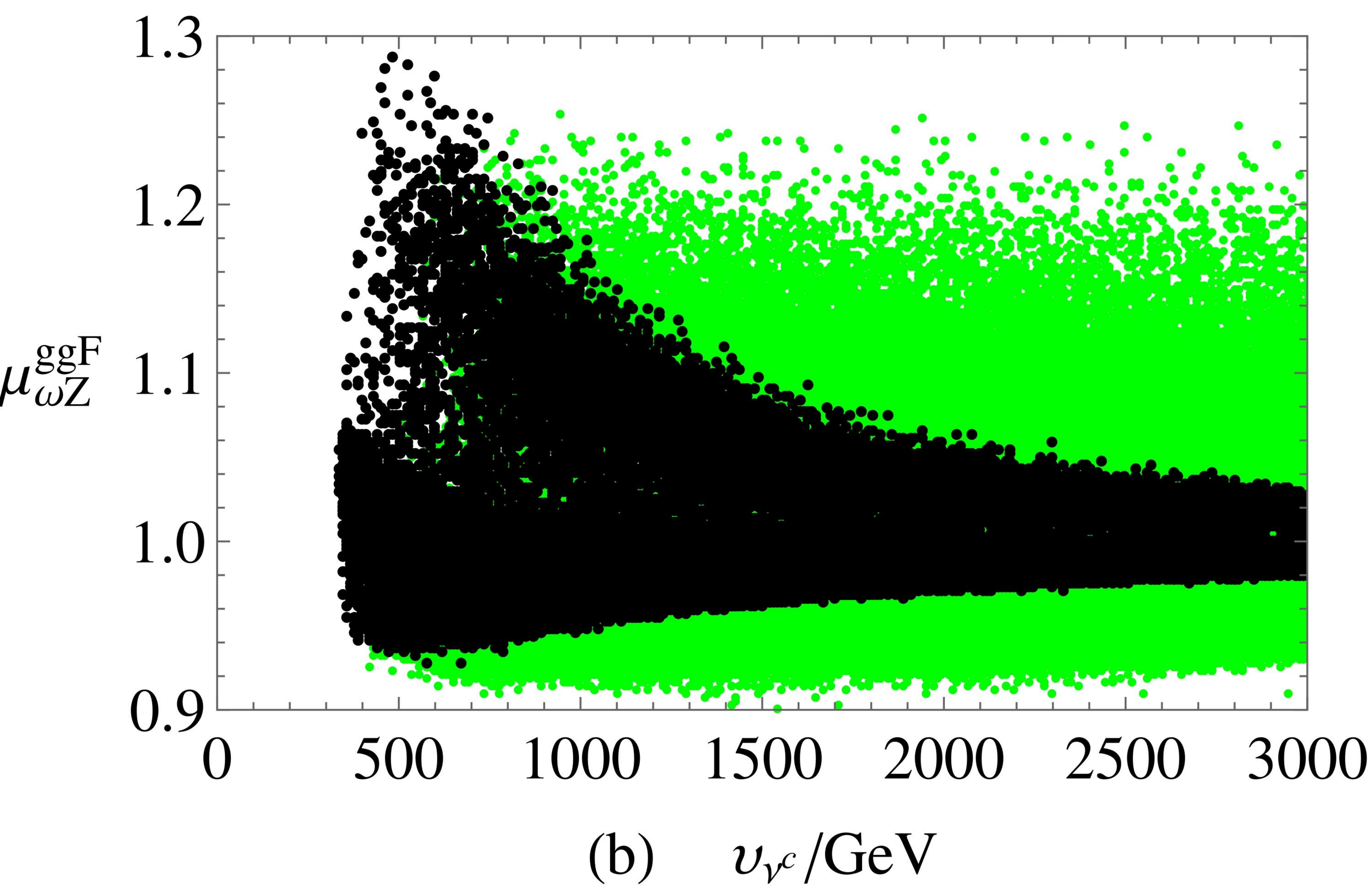}
\end{minipage}%
\\
\begin{minipage}[c]{0.5\textwidth}
\centering
\includegraphics[width=3in]{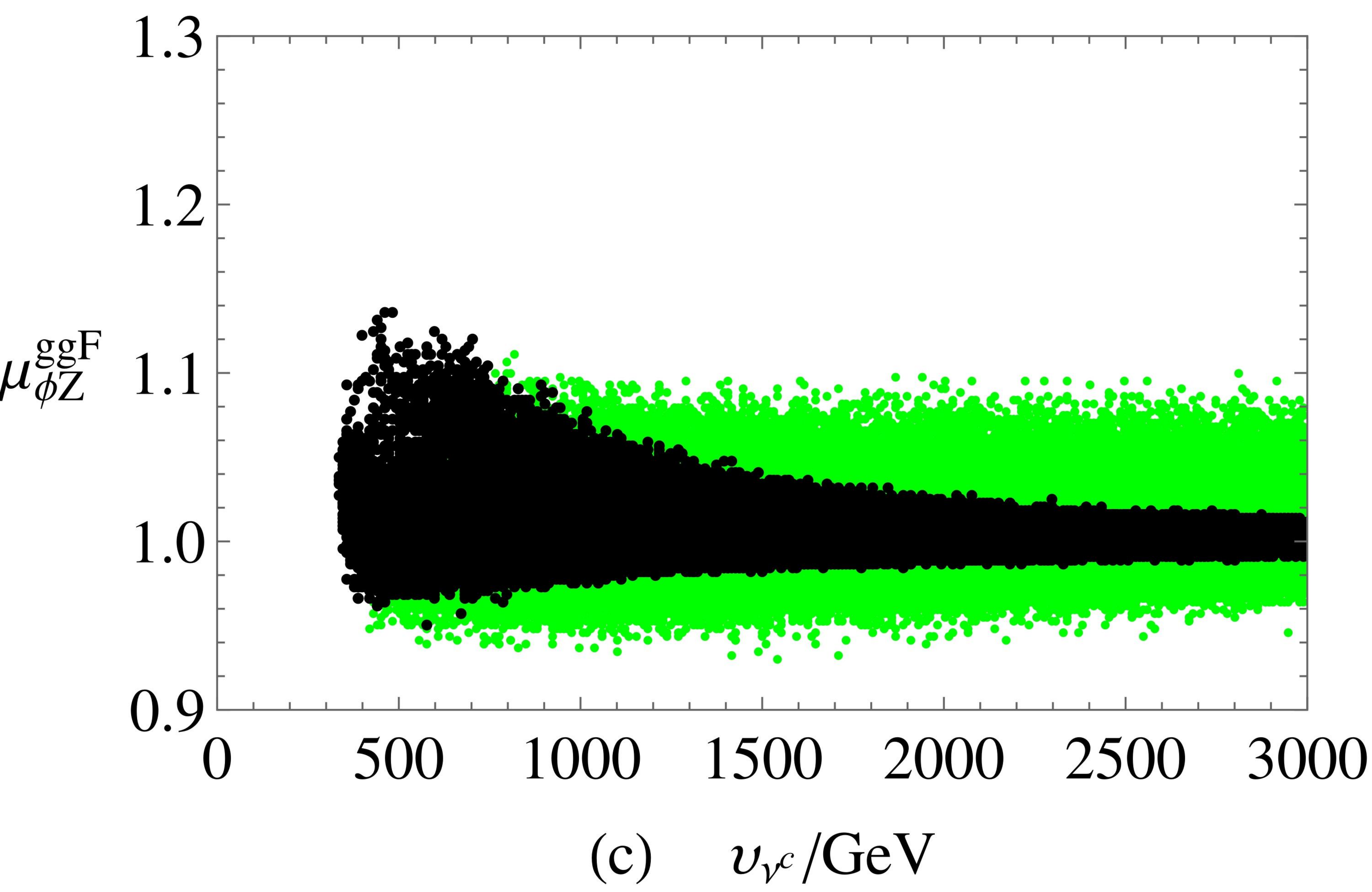}
\end{minipage}%
\begin{minipage}[c]{0.5\textwidth}
\centering
\includegraphics[width=3in]{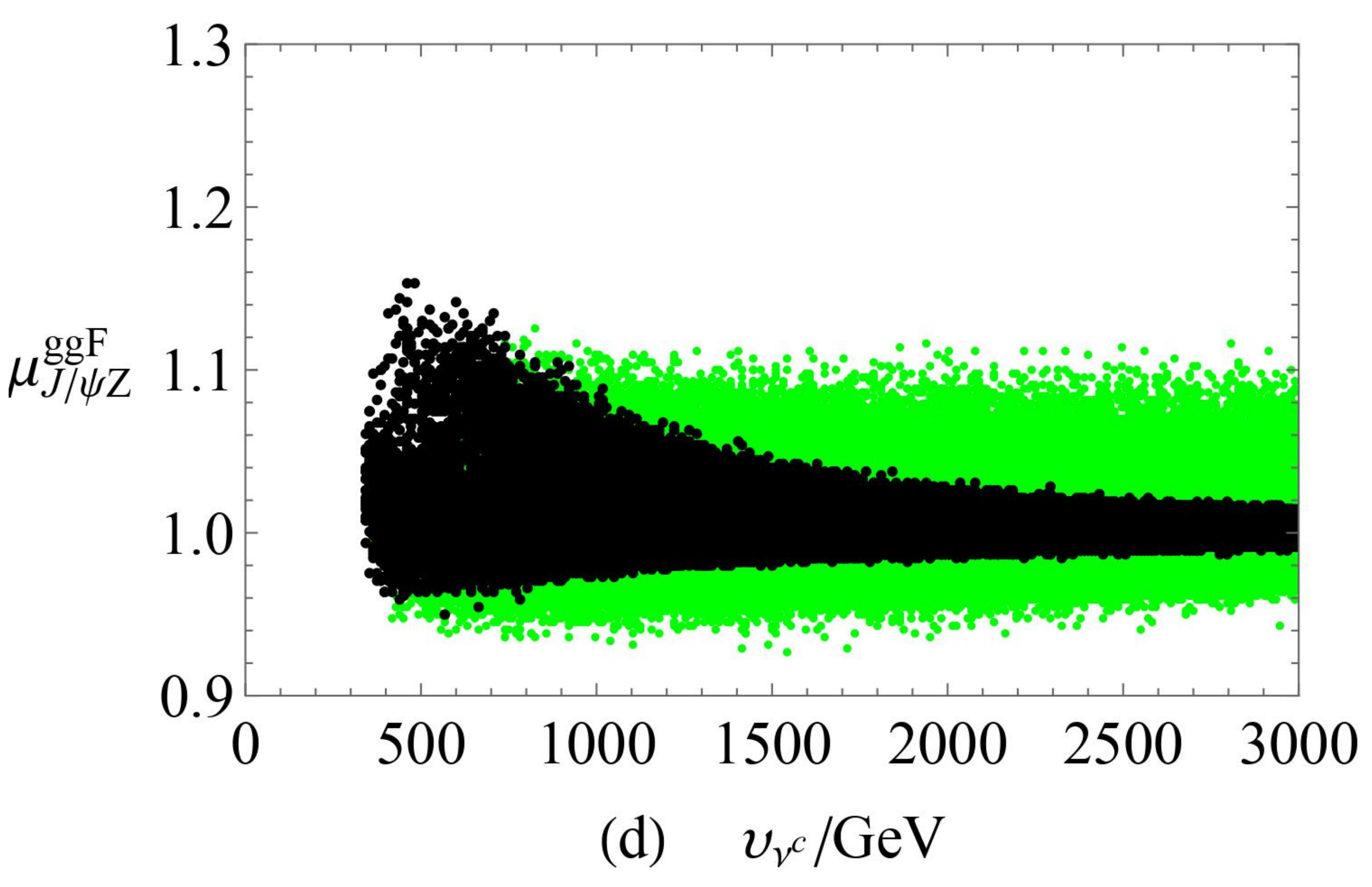}
\end{minipage}%
\\
\begin{minipage}[c]{0.5\textwidth}
\centering
\includegraphics[width=3in]{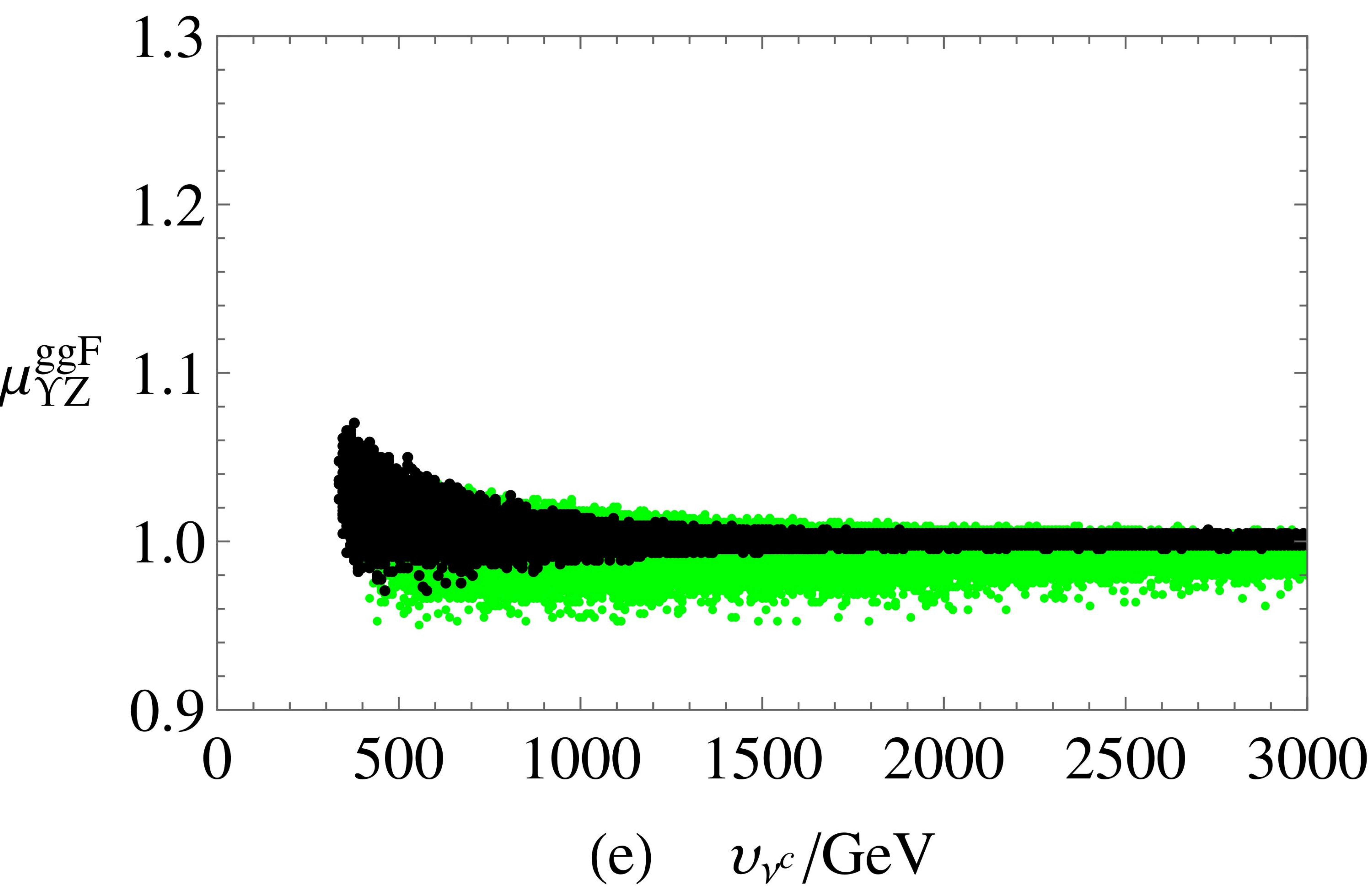}
\end{minipage}%
\begin{minipage}[c]{0.5\textwidth}
\centering
\includegraphics[width=3in]{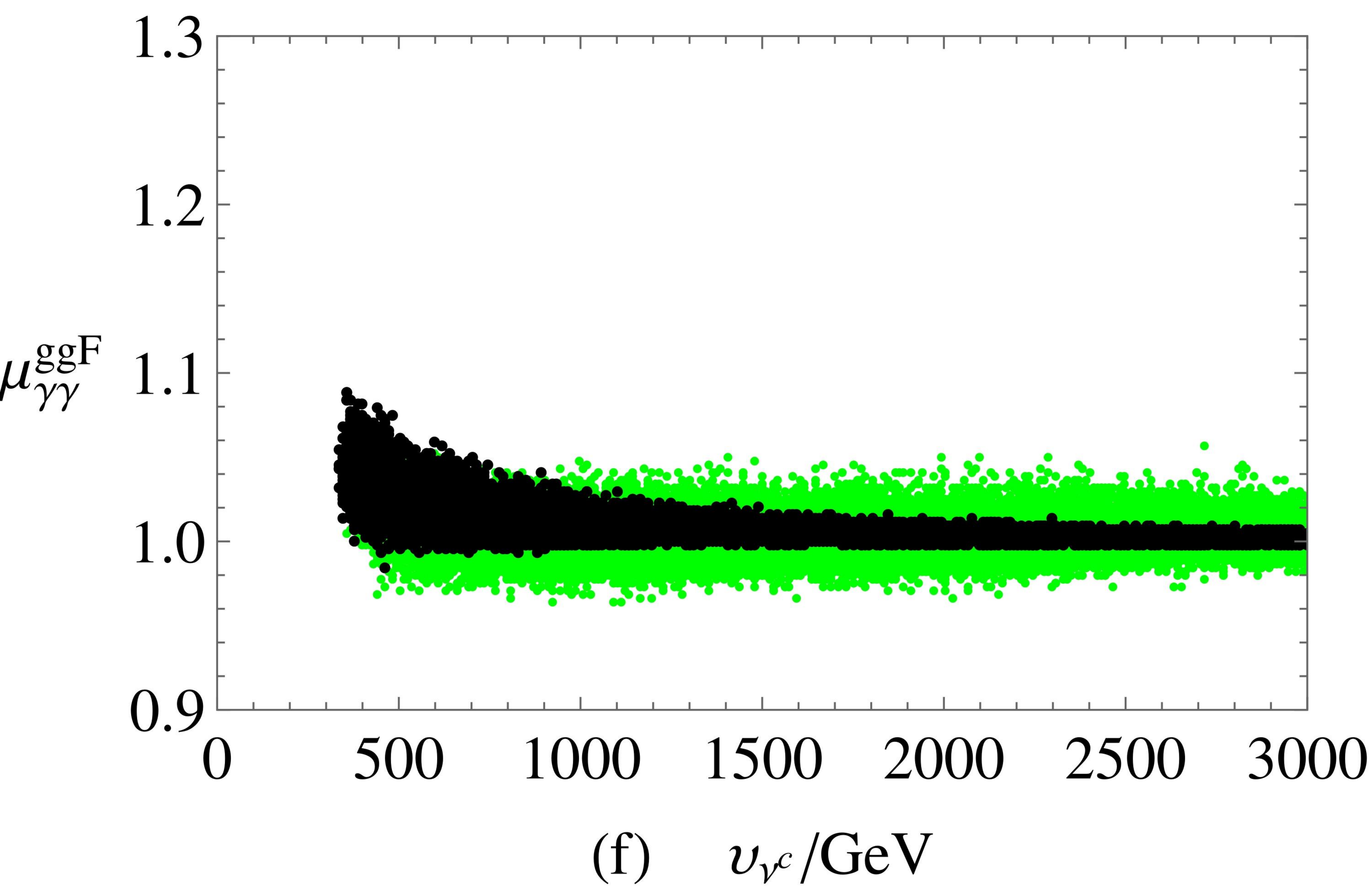}
\end{minipage}%
\caption[]{The signal strengths $\mu_{MZ,\gamma\gamma}^{\rm{ggF}}$ versus the parameter $\upsilon_{\nu^{c}}$. The green dots represent $0.02\leq\lambda<0.1$, and black dots represent $0.1\leq\lambda\leq0.2$.}
\label{h1-h2}
\end{figure}

Then, through a random scan in Table \ref{tabscan} according to the above constraints, we plot the signal strengths $\mu_{MZ,\gamma\gamma}^{\rm{ggF}}$ varying with the parameter $\upsilon_{\nu^{c}}$ in Fig. \ref{h1-h2}. The dots in the figures are the corresponding physical quantity's values of the remaining parameters after being constrained by the above discussions. We divide the  parameter $\lambda$ into two scanning ranges shown in the figures, that is $0.02\leq\lambda<0.1$ (green dots) and $0.1\leq\lambda\leq0.2$ (black dots) to see the correlation with the parameter $\lambda$ for the results.

When $\tan\beta=10,\lambda<0.1$, the lightest Higgs mass in the tree-level in the $\mu\nu$SSM is close to that in the MSSM. But when $\tan\beta=10,\lambda>0.1$, in Fig. \ref{mh} (a) the lightest Higgs boson mass drops down quickly, and the mixing of the neutral components of the Higgs doublets with the right-handed sneutrinos will induce that the Higgs couplings in the $\mu\nu$SSM are more different than those in the MSSM.
In Eq.~(\ref{muss}), we can know that the signal strengths $\mu_{MZ}^{\rm{ggF}}$ are obviously affected by ${{\Gamma_{\mu\nu\rm{SSM}}(h\rightarrow MZ)}\over{\Gamma_{\rm{SM}}(h\rightarrow MZ)}}$. To help to analyze the results, we firstly define the ratio
\begin{eqnarray}
R_{M}={{\Gamma_{\mu\nu\rm{SSM}}(h\rightarrow MZ)}\over{\Gamma_{\rm{SM}}(h\rightarrow MZ)}}.
\end{eqnarray}

In Fig. \ref{h1-h2}, the numerical results show that the signal strengths $\mu_{\rho Z}^{\rm{ggF}}$ (a) and $\mu_{\omega Z}^{\rm{ggF}}$ (b) in the $\mu\nu$SSM can easily reach 1.2,  which display a large deviation from the SM. One can also see that the signal strengths $\mu_{\rho Z}^{\rm{ggF}}$ (a) and $\mu_{\omega Z}^{\rm{ggF}}$ (b) are similar. In Eq.~(\ref{form-factors1}) and Eq.~(\ref{form-factors2}), we can know that the indirect contributions $F_{||\,ind}^{MZ}$ and $F_{\perp\,ind}^{MZ}$ for the decay width of $h\rightarrow MZ$ have two terms, where the first term ${{\kappa_{Z}}\over{1-r_{M}/r_{Z}}}\sum\limits_{q}f_{M}^q\upsilon_{q}$ is not related to NP which only related to the meson decay constants and meson masses, and the NP contributions are reflected in the second term. For the decay width of $h\rightarrow MZ$ in Eq.~(\ref{decay-width}), we can see that meson masses have a great influence on ${{8r_{M}r_{Z}}\over{(1-r_{Z}-r_{M})^2}}$, which can affect the contributions from $|F_{\perp\,ind}^{MZ}|^2$ compared to  $|F_{||\,ind}^{MZ}|^2$. Therefore, the signal strengths $\mu_{M Z}^{\rm{ggF}}$ are affected by the NP contributions,  the meson decay constants and meson masses. To display explicitly, we  calculate the indirect contributions for the decay width of $h\rightarrow MZ$ shown in Table \ref{tabmeson}. Here, $C_{\gamma Z}=C_{\gamma Z}^{SM}+C_{\gamma Z}^{NP}$, with $C_{\gamma Z}^{SM}\simeq-2.43$. Through Table \ref{tabmeson}, we can clearly see that $R_{\rho}-1 \simeq 0.135\times(|C_{\gamma Z}^{NP}-2.37|^2-2.37^2)$ is similar to $R_{\omega}-1 \simeq 0.137\times(|C_{\gamma Z}^{NP}-2.47|^2-2.47^2)$, which can indicate that the signal strength $\mu_{\rho Z}^{\rm{ggF}}$ is similar to $\mu_{\omega Z}^{\rm{ggF}}$.

\begin{table}
\begin{tabular}{| c | c | c | c | c | c |}
\hline
  \quad & \bm{${\rho}$} & \bm{$\omega$} & \bm{$\phi$} & \rm{J}/\bm{$\psi$} & \bm{$\Upsilon$} \\
  \hline
  \bm{$F_{\parallel ind,NP}^{MZ}$} & $\makecell[c]{0.042\\+4.3\times10^{-4}C_{\gamma Z}}$ & $\makecell[c]{-0.01\\+1.3\times10^{-4}C_{\gamma Z}}$ & $\makecell[c]{-0.039\\-2\times10^{-4}C_{\gamma Z}}$ & $\makecell[c]{0.04\\+7.5\times10^{-4}C_{\gamma Z}}$ & $\makecell[c]{-0.12\\-6.5\times10^{-4}C_{\gamma Z}}$ \\
  \hline
  \bm{$F_{\parallel ind,SM}^{MZ}$} & $\makecell[c]{0.041}$ & $\makecell[c]{-0.01}$ & $\makecell[c]{-0.039}$ & $\makecell[c]{0.04}$ & $\makecell[c]{-0.12}$ \\
  \hline
  \bm{$F_{\perp ind,NP}^{MZ}$} & $\makecell[c]{0.042\\+1.18C_{\gamma Z}}$ & $\makecell[c]{-0.01\\+0.34C_{\gamma Z}}$ & $\makecell[c]{-0.039\\-0.33C_{\gamma Z}}$ & $\makecell[c]{0.04\\+0.13C_{\gamma Z}}$ & $\makecell[c]{-0.12\\-0.012C_{\gamma Z}}$ \\
  \hline
  \bm{$F_{\perp ind,SM}^{MZ}$} & $-2.83$ & $-0.84$ & $0.76$ & $-0.27$ & $-0.093$ \\
  \hline
  \bm{${{8r_{M}r_{Z}}\over{(1-r_{Z}-r_{M})^2}}$} & $7.3\times10^{-4}$ & $7.5\times10^{-4}$ & $1.3\times10^{-3}$ & $1.2\times10^{-2}$ & 0.11 \\
  \hline
  \bm{$R_{M}-1$} & $\makecell[c]{0.135\times\\(|C_{\gamma Z}^{NP}-2.37|^2\\-2.37^2)}$ & $\makecell[c]{0.137\times\\(|C_{\gamma Z}^{NP}-2.47|^2\\-2.47^2)}$ & $\makecell[c]{0.0614\times\\(|C_{\gamma Z}^{NP}-2.25|^2\\-2.25^2)}$ & $\makecell[c]{0.0813\times\\(|C_{\gamma Z}^{NP}-1.95|^2\\-1.95^2)}$ & $\makecell[c]{0.001\times\\(|C_{\gamma Z}^{NP}+12.9|^2\\-12.9^2)}$\\
  \hline
\end{tabular}
\caption{The indirect contributions for the decay width of $h\rightarrow MZ$, where $C_{\gamma Z}=C_{\gamma Z}^{SM}+C_{\gamma Z}^{NP}$, with $C_{\gamma Z}^{SM}\simeq-2.43$. }
\label{tabmeson}
\end{table}

The numerical results in Fig. \ref{h1-h2} show that the signal strengths $\mu_{\phi Z}^{\rm{ggF}}$ (c) and $\mu_{J/\psi Z}^{\rm{ggF}}$ (d) in the $\mu\nu$SSM can reach about 1.1, which display a deviation about 10\% from the SM. The decays $h\rightarrow\phi Z$ and $h\rightarrow J/\psi Z$ in the $\mu\nu$SSM may be more easily detected than those in the SM, if the future colliders can detect them.

Here in Table \ref{tabmeson}, $F_{\parallel ind,NP}^{MZ}$ and $F_{\perp ind,NP}^{MZ}$ represent the indirect longitudinal and transverse form factors for NP, $F_{\parallel ind,SM}^{MZ}$ and $F_{\perp ind,SM}^{MZ}$ represent these for the standard model, and $M$ is the vector meson, respectively. For the mesons $\phi$ and $J/\psi$, $|F_{\parallel ind,NP}^{\phi Z,J/\psi Z}|^2$ will exceed ${{8r_{M}r_{Z}}\over{(1-r_{Z}-r_{M})^2}}|F_{\perp ind,NP}^{\phi Z,J/\psi Z}|^2$, but the ratio $|F_{\parallel ind,NP}^{\phi Z,J/\psi Z}|^2/|F_{\parallel ind,SM}^{\phi Z,J/\psi Z}|^2$ will be still less than $|F_{\perp ind,NP}^{\phi Z,J/\psi Z}|^2/|F_{\perp ind,SM}^{\phi Z,J/\psi Z}|^2$. Then, one can know that $|F_{\parallel ind,NP}^{\phi Z}|^2$ and $|F_{\parallel ind,NP}^{J/\psi Z}|^2$ have a large impact on the decay width in Eq.~(\ref{decay-width}), but they have less impact on NP than $|F_{\perp ind,NP}^{\phi Z}|^2$ and $|F_{\perp ind,NP}^{J/\psi Z}|^2$. Therefore $|F_{\perp ind,NP}^{\phi Z}|^2$ and $|F_{\perp ind,NP}^{J/\psi Z}|^2$ are still the main factors affecting $R_{\rho}$ and $R_{J/\psi}$. From Table \ref{tabmeson}, we can know $F_{\perp ind,NP}^{\phi Z}=-0.039-0.33C_{\gamma Z}$ and $F_{\perp ind,NP}^{J/\psi Z}=0.04+0.13C_{\gamma Z}$, which the NP contributions will be slightly weakened by ${{\kappa_{Z}}\over{1-r_{M}/r_{Z}}}\sum\limits_{q}f_{M}^q\upsilon_{q}$ in Eq.~(\ref{form-factors2}). But for the NP contributions, the ratio $|F_{\perp ind,NP}^{J/\psi Z}|^2/|F_{\perp ind,SM}^{J/\psi Z}|^2$ is approximate to $|F_{\perp ind,NP}^{\phi Z}|^2/|F_{\perp ind,SM}^{\phi Z}|^2$. In Table \ref{tabmeson}, one can clearly see that $R_{\phi}-1 \simeq 0.0614\times(|C_{\gamma Z}^{NP}-2.25|^2-2.25^2)$ is similar to $R_{J/\psi }-1 \simeq 0.0813\times(|C_{\gamma Z}^{NP}-1.95|^2-1.95^2)$. Therefore,  the signal strengths $\mu_{\phi Z}^{\rm{ggF}}$ and $\mu_{J/\psi Z}^{\rm{ggF}}$ are similar.

In Fig. \ref{h1-h2}(e), the signal strength $\mu_{{{\Upsilon} Z}}^{\rm{ggF}}$  in the $\mu\nu$SSM is close to 1, which means that  is difficult to find NP through the signal strength $\mu_{{{\Upsilon} Z}}^{\rm{ggF}}$.
In the five decays $h\rightarrow M Z$ here, the decay width $\Gamma{{(h\rightarrow \Upsilon Z)}}$ is the largest one, which the associated decay constant $f_\Upsilon$ and meson mass $m_\Upsilon$ are the largest one. Normalized to the SM expectation, the terms related to the SM part will be counteracted. And the ratio $R_{\Upsilon}={{\Gamma_{\mu\nu\rm{SSM}}(h\rightarrow \Upsilon Z)}\over{\Gamma_{\rm{SM}}(h\rightarrow \Upsilon Z)}}$ can be seen how large deviation relative to the SM  expectation of the decay width for $h\rightarrow \Upsilon Z$.
Through  Table ~\ref{tabmeson}, one can find that $F_{\perp ind,NP}^{\Upsilon Z}=-0.12-0.012C_{\gamma Z}$, which the first term is ten times bigger than the coefficient of $C_{\gamma Z}$. Therefore, the first term of $F_{\perp ind,NP}^{\Upsilon Z}$  including the associated meson mass and decay constant can weak the NP contributions to the ratio $R_{\Upsilon}$. Then,  one can clearly see that $R_{\Upsilon}-1 \simeq 0.001\times(|C_{\gamma Z}^{NP}+12.9|^2-12.9^2)$ in Table ~\ref{tabmeson}, where the coefficient is 0.001 which makes the NP contributions $C_{\gamma Z}^{NP}$ difficult to improve the signal strength $\mu_{{{\Upsilon} Z}}^{\rm{ggF}}\approx {{\Gamma_{\rm{SM}}^{h}}\over{\Gamma_{\rm{NP}}^{h}}}{{\Gamma_{\rm{NP}}(h\rightarrow gg)}\over{\Gamma_{\rm{SM}}(h\rightarrow gg)}}R_{\Upsilon}$.

For the famous decay channel $h\rightarrow\gamma\gamma$, we also show the signal strength $\mu_{\gamma\gamma}^{\rm{ggF}}$ in Fig. \ref{h1-h2}(f). The numerical results show that the signal strength $\mu_{\gamma\gamma}^{\rm{ggF}}$  in the $\mu\nu$SSM can also reach about 1.08, which display a deviation about 8\% from the SM. Fig. \ref{h1-h2}(f) shows that the signal strength $\mu_{\gamma\gamma}^{\rm{ggF}}\geq0.96$, but the experimental value of $\mu_{\gamma\gamma}^{\rm{ggF}}$ in Eq. (\ref{hdecay}) is $\mu_{\gamma\gamma}^{\rm{ggF}}\geq0.93$ at $2\sigma$ level. That is to say, our results are experimentally acceptable.

In Fig. \ref{h1-h2} (a-d), during the calculation of random scanning, the coupling $g^{L}_{h\widetilde{\chi}^{\pm}_{1}\widetilde{\chi}^{\mp}_{1}}$ will be more strong as $M_2\approx\mu$, and then the signal strengths in Fig. \ref{h1-h2} (a-d)  will be large. When $M_2\neq\mu$, the coupling will be weak, and then the signal strengths will be small.

Through Fig. \ref{h1-h2},  we can see that black dots ($0.1\leq\lambda\leq0.2$) are all close to 1, when $\upsilon_{\nu^{c}}>1500$ GeV. Because the parameters $\lambda$ and $\upsilon_{\nu^{c}}$ affect the parameter $\mu=3\lambda\upsilon_{\nu^{c}}$, which can change the lighter chargino mass. When $0.1\leq\lambda$ and $\upsilon_{\nu^{c}}>1500$ GeV, the  lighter chargino mass is heavier than about $450$ GeV. Here the lighter chargino mass can affect the signal strengths $\mu_{MZ,\gamma\gamma}^{\rm{ggF}}$ obviously. And then we can know that when the lighter chargino mass is heavier than about $450$ GeV, the NP contributions for the signal strengths $\mu_{MZ,\gamma\gamma}^{\rm{ggF}}$ will be small.

\begin{figure}
\setlength{\unitlength}{1mm}
\begin{minipage}[c]{0.5\textwidth}
\centering
\includegraphics[width=3in]{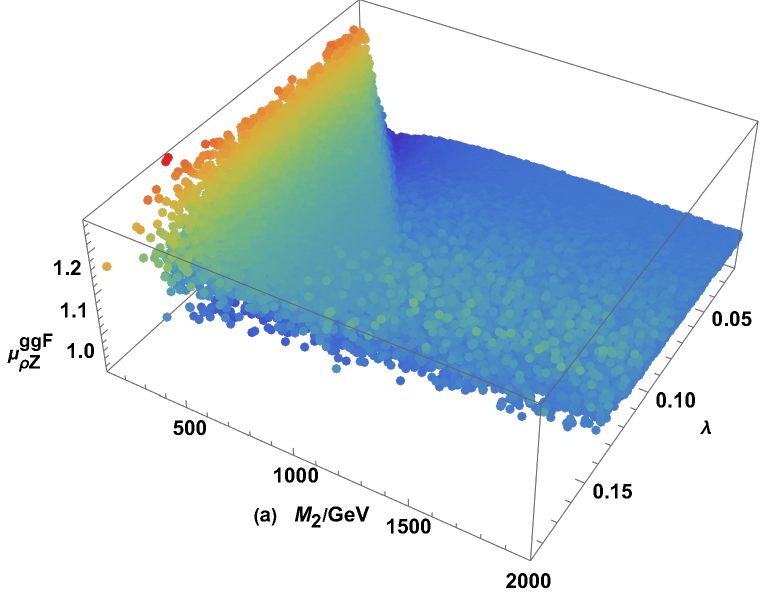}
\end{minipage}%
\begin{minipage}[c]{0.5\textwidth}
\centering
\includegraphics[width=3in]{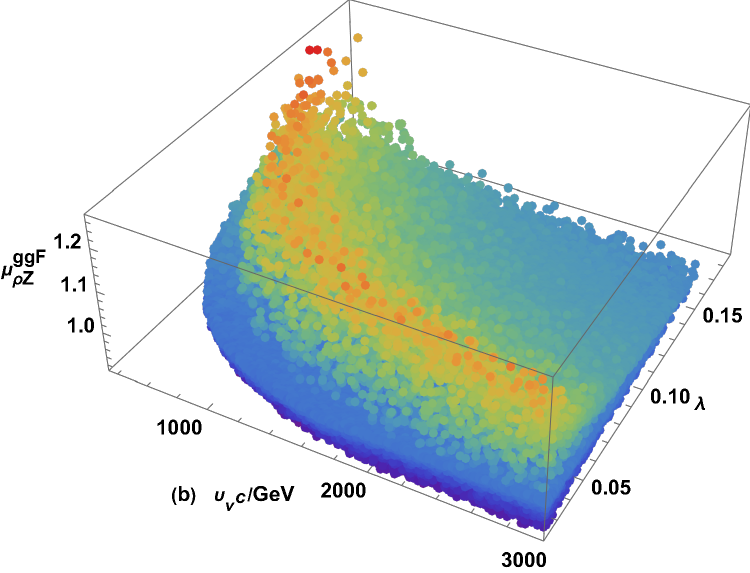}
\end{minipage}%
\caption[]{The 3-d plots of the signal strength $\mu_{\rho Z}^{\rm{ggF}}$ depend on two parameters $\lambda$ and $M_2$ (a), or $\lambda$ and $\upsilon_{\nu^{c}}$ (b).}
\label{3D}
\end{figure}

The 3-d plots can help us realize how the signal strengths depend on these parameters $M_{2}$, $\lambda, \upsilon_{\nu^c}$. Fig. \ref{3D} (a) shows a 3-d plot where the signal strength $\mu_{\rho Z}^{ggF}$ versus the two parameters $\lambda$ and $M_{2}$. To show the signal strength $\mu_{\rho Z}^{ggF}$ in gradients, we picture with the different colour in gradients. These orange and yellow points represent large $\mu_{\rho Z}^{ggF}$ and blue points represent small $\mu_{\rho Z}^{ggF}$. For small $M_{2}$, we can see that the signal strength $\mu_{\rho Z}^{ggF}$ can reach 1.2, meanwhile the parameter $\lambda$ can be in the range of 0.02 to 0.18.  When $M_{2}>$ 500 GeV, the signal strength $\mu_{\rho Z}^{ggF}$ is close to 1, meanwhile the parameter $\lambda$ is in the range of 0.02 to 0.14. We also picture a 3-d plot for the signal strength $\mu_{\rho Z}^{ggF}$ versus the two parameters $\lambda$ and $\upsilon_{\nu^{c}}$ in Fig. \ref{3D} (b). The numerical results shows that the signal strength $\mu_{\rho Z}^{ggF}$ will be very small in large $\lambda$ and large $\upsilon_{\nu^{c}}$.

\begin{figure}
\setlength{\unitlength}{1mm}
\begin{minipage}[c]{0.7\textwidth}
\includegraphics[width=4.5in]{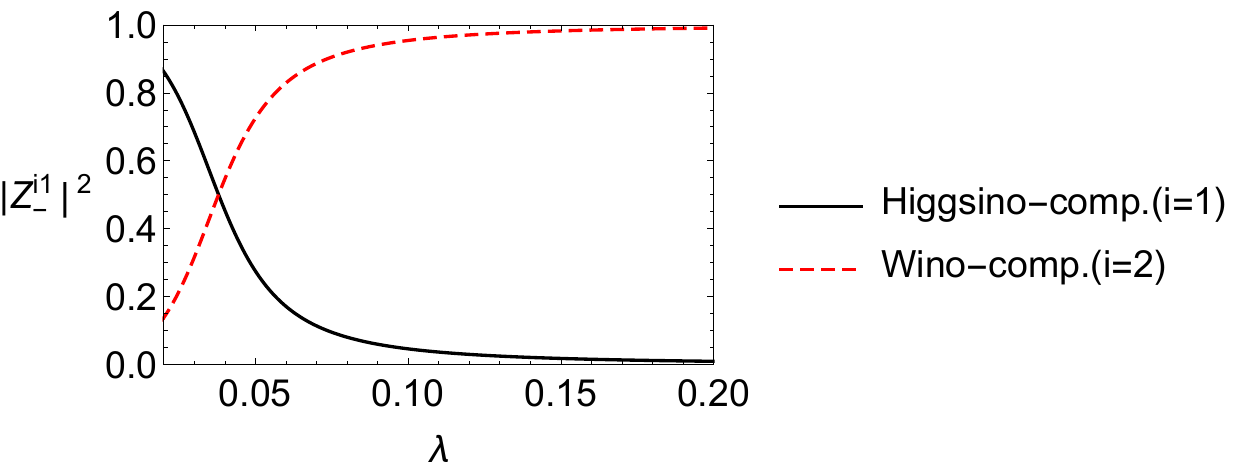}
\end{minipage}%
\caption[]{Higgsino-component and Wino-component of the lighter chargino $\tilde{\chi}_{1}^{\pm}$ versus the parameter $\lambda$, which are defined by $|Z_{-}^{i1}|^2$ with $i$=1,2.}
\label{HW}
\end{figure}

To be more clarity, we plot Higgsino-component and Wino-component of the lighter chargino $\tilde{\chi}_{1}^{\pm}$ versus $\lambda$ in Fig. \ref{HW}, when $M_{2}=200$ GeV, $\tan\beta=5$, $\upsilon_{\nu^c}=2000$ GeV, $\kappa=0.5$. One can find that when $\lambda=0.02$, namely $M_{2}>\mu$, the main component of the $\tilde{\chi}_{1}^{\pm}$ is the Higgsino. when $\lambda>0.05$, namely $M_{2}<\mu$, the main component of the $\tilde{\chi}_{1}^{\pm}$ is the Wino. So, for small $M_{2}$, such as $M_{2}<200$ GeV,  the lighter chargino $\tilde{\chi}_{1}^{\pm}$ is mainly Wino-like state in Fig. 9, when $M_{2}<\mu$.

\section{Conclusion\label{sec5} }

In this paper, we have discussed the lightest SM-like Higgs boson rare decays $h\rightarrow MZ$ with the meson $M=\rho, \omega, \phi, J/\psi, \Upsilon$ in the framework of the $\mu\nu$SSM. In the $\mu\nu$SSM, the left- \& right-handed sneutrino VEVs lead to the mixing of the neutral components of the Higgs doublets with the left- \& right-handed sneutrinos, which can give a rich phenomenology in the Higgs sector of  the $\mu\nu$SSM. The mixing of the neutral components of the Higgs doublets with the right-handed sneutrinos affect the mass and the coupling of the lightest Higgs boson, which can give new effect to the lightest Higgs boson weak hadronic decays $h\rightarrow MZ$.

The numerical results show that the NP contributions to the processes $h\rightarrow \rho Z$ and $h\rightarrow \omega Z$ are more considerable. When chargino mass is small, the signal strengths $\mu_{\rho Z,\omega Z}^{\rm{ggF}}$ in the $\mu\nu$SSM can reach 1.2, that display a large deviation from the SM and is experimentally promising to find NP. For decays $h\rightarrow MZ$ ($M=\phi, J/\psi$), the signal strengths $\mu_{\phi Z,J/\psi Z}^{\rm{ggF}}$  in the $\mu\nu$SSM can attain about 1.1, which display a deviation about 10\% from the SM. The decays $h\rightarrow MZ$ may be accessible at a potential 100 TeV collider or the other future high energy colliders \cite{100TeV}.

\begin{acknowledgments}
\indent\indent
The work has been supported by the National Natural Science Foundation of China (NNSFC) with Grants No. 11705045, No. 12075074, No. 12235008, No. 11535002, Natural Science Foundation for Distinguished Young Scholars of Hebei Province with Grant No. A2022201017, Natural Science Foundation of Guangxi Autonomous Region with Grant No. 2022GXNSFDA035068, the youth top-notch talent support program of the Hebei Province, and Midwest Universities Comprehensive Strength Promotion project.
\end{acknowledgments}

\end{document}